\documentclass[fleqn,usenatbib]{mnras}
\usepackage{newtxtext,newtxmath}
\usepackage[T1]{fontenc}
\DeclareRobustCommand{\VAN}[3]{#2}
\let\VANthebibliography\thebibliography
\def\thebibliography{\DeclareRobustCommand{\VAN}[3]{##3}\VANthebibliography}
\usepackage{graphicx}	% Including figure files
\usepackage{amsmath}	% Advanced maths commands
\usepackage{physics}

\title[Edge-on debris disk fitting]{\texttt{Rave}: A non-parametric method for recovering the surface brightness and height profiles of edge-on debris disks}

\author[Han et al.]{
Y~Han,$^{1}$\thanks{E-mail: yinuo.han@ast.cam.ac.uk}
M~C~Wyatt,$^{1}$
L~Matr{\`a},$^{2, 3}$
\\
$^{1}$Institute of Astronomy, University of Cambridge, Madingley Road, Cambridge CB3 0HA, UK\\
$^{2}$School of Physics, National University of Ireland Galway, University Road, Galway, Ireland\\
$^{3}$School of Physics, Trinity College Dublin, The University of Dublin, College Green, Dublin 2, Ireland 
}

\date{Accepted 2022 February 8. Received 2022 February 8; in original form 2021 November 15}

\pubyear{2021}

\begin{document}
\label{firstpage}
\pagerange{\pageref{firstpage}--\pageref{lastpage}}
\maketitle

\begin{abstract}
Extrasolar analogues of the Solar System's Kuiper belt offer unique constraints on outer planetary system architecture. Radial features such as the sharpness of disk edges and substructures such as gaps may be indicative of embedded planets within a disk. Vertically, the height of a disk can constrain the mass of embedded bodies. 
Edge-on debris disks offer a unique opportunity to simultaneously access the radial and vertical distribution of material, however recovering either distribution in an unbiased way is challenging. 
In this study, we present a non-parametric method to recover the surface brightness profile (face-on surface brightness as a function of radius) and height profile (scale height as a function of radius) of azimuthally symmetric, edge-on debris disks. The method is primarily designed for observations at thermal emission wavelengths, but is also applicable to scattered light observations under the assumption of isotropic scattering. 
By removing assumptions on underlying functional forms, this algorithm provides more realistic constraints on disk structures. 
We also apply this technique to ALMA observations of the AU~Mic debris disk and derive a surface brightness profile consistent with estimates from parametric approaches, but with a more realistic range of possible models that is independent of parametrisation assumptions. Our results are consistent with a uniform scale height of 0.8~au, but a scale height that increases linearly with radius is also possible. 
\end{abstract}


%%%%%%%%%%%%%%%%% BODY OF PAPER %%%%%%%%%%%%%%%%%%

\section{Introduction}
Circumstellar disks of planetesimals, asteroids, dust and gas are a prevalent feature of planetary systems \citep{Sibthorpe2018}. They are considered to be analogues of the asteroid and Kuiper belts of the Solar System, and similarly carry signatures of dynamical interaction with planets in the system. 

Material in these exo-Kuiper belts --- or ``debris disks’’ --- evolve under a collisional cascade in which collisions gradually grind down planetesimals into smaller and smaller grains \citep{Wyatt2008}. While the mass of a debris disk is concentrated in the largest bodies (e.g., planetesimals), most of the observed flux originates from small dust grains near the bottom of the collisional cascade which collectively possess the bulk of the surface area. These dust disks are often observed through scattered light at optical to near-infrared wavelengths or through thermal emission at mid-infrared to mm wavelengths. 

% $\mu$m-sized grains which dominate the $\mu$m flux are near the bottom of the collisional cascade and may be readily displaced from the parent planetesimal belt by stellar radiation pressure. Mm-sized dust, however, are significantly less affected by radiation pressure and are expected to closely trace the distribution of the parent belt consisting of planetesimals.

Observations have revealed that substructures such as gaps and asymmetries may be common in debris disks \citep{Marino2020}, and numerous studies have linked substructures to dynamical interactions with planets \citep{Krivov2010, Matthews2014, Hughes2018b}. 
% Debris disks are continuously shaped by both planet-disk interactions and the disk’s self interaction as the system continues to evolve. Planet-disk interactions such as secular perturbation, resonance and scattering may alter the orbits of material in the debris disk and thus the disk’s structure. 
Observational examples include a warped geometry in the Beta Pic debris disk \citep{Golimowski2006} which is hypothesised to be due to gravitational perturbations from an inclined planet \citep{Mouillet1997}. A planet satisfying the required properties to form the warp, Beta Pic b, was subsequently detected through direct imaging \citep{Lagrange2009}. Another example is the HR8799 system which hosts four directly-imaged planets \citep{Marois2008, Marois2010}. The debris disk in this system is thought to comprise of a scattered disk of comets in addition to the low-eccentricity primary disk, which may have been shaped by planet migration in a way similar to the Solar System's Kuiper belt \citep{Geiler2019}.

Debris disks provide the best constraints of planetary architecture in the outer planetary system, where planets are often difficult to detect. Recent planet imaging surveys such as the SHINE survey with VLT/SPHERE \citep{Vigan2021} and the GPIES survey with Gemini/GPI \citep{Nielsen2019} have increased the sample of outer planets being detected. These detections provide opportunities to better confirm the link between disk substructures and planets, raising the prospect of directly using debris disk images to detect exoplanets. 

% It is therefore reasonable to say that debris disks serve as records of the history of planetary systems, and their structure and composition amplify planetary dynamics that are otherwise difficult to observe. 
The close link between planets and disks has motivated much attention to observe and theoretically model detailed disk structures. However, gaining a full picture of debris disks in three-dimensional space is difficult. For disks at low inclinations (i.e., close to face-on), their face-on flux distribution is well characterised, but their vertical height is hardly constrained.
%Scale heights are important for estimating the level of dynamical stirring within the disk, which in turn constrains the presence of large embedded bodies within the disk \citep{Daley2019}. 
%\citet{Daley2019}, for example, measured an H/r of 0.031 in the AU~Mic debris disk with ALMA images, which was used to place an upper limit of 1.8 Earth masses of the largest stirring body, ruling out perturbations due to Neptune analogues or larger in the outer disk.

On the other hand, edge-on disks offer unique opportunities to study their vertical height distributions. However, high inclinations compromise access to the face-on radial surface brightness profile. For an edge-on disk, the observed flux at each point along the disk has contributions from material physically located at a range of different radii from the central star. Disentangling the projected, point spread function (PSF)-convolved flux to recover the face-on surface brightness distribution is therefore not a trivial problem. 

This study focuses on the case of nearly edge-on disks (i.e., highly inclined). Several studies have attempted to recover the face-on radial surface brightness profile of edge-on disks and fit their scale heights, most of which taking a parametric approach, i.e. assuming a functional form for the radial profile and fitting to its functional parameters. Models for the radial profile invoked using this approach include a Gaussian (Beta Pic, \citealp{Matra2019}), top-hat + star-centred peak (AU~Mic, \citealp{MacGregor2013}), power law (49~Ceti, \citealp{Hughes2017}), a combination of a rising and falling power law (HD32297 and HD61005, \citealp{MacGregor2018}) and two rings separated by a Gaussian gap (HD15115, \citealp{MacGregor2019}).
%\citet{Hughes2017} fitted ALMA images of the debris disk of 49 Ceti with a single power law, double power law and a single power law with an additional ring. 
%Upon realising that a Gaussian model does not offer a reasonable fit, \citet{MacGregor2018} fitted the debris disks of HD32297 and HD61005 with an inner parent body belt and an outer halo, which is effectively a rising power law followed by a falling one. 
%\citet{MacGregor2019} modelled the debris disk of HD15115 with a two-ring model and a ring with a Gaussian gap model. 

Although parametric fitting is able to generate reasonable matches to the observed image in most cases, the derived uncertainties on the fitted parameters and the resulting profile are only meaningful if the disk profile indeed has the assumed functional form. 
%The uncertainties on the resulting fitted profile is therefore only known given the assumed functional form. 
Such fitting may also bias the recovered radial profile towards the assumed functional form, potentially obscuring or artificially generating substructures which are crucial for subsequently interpreting the disk’s structural and dynamical properties. 

To mitigate such model-dependent biases, an attempt has been made to recover the surface brightness profile non-parametrically in the context of the Beta Pic debris disk. \citet{Telesco2005} developed a method to fit to the radial surface brightness profile while assuming that the disk is azimuthally symmetric, which models the disk as consisting of a series of concentric annuli, each with uniform density and temperature. The relative brightness of each annulus was iteratively modified until the resulting flux profile of the edge-on model image reproduced that of the observation. This method was applied to both mid-IR \citep{Telesco2005} and ALMA observations \citep{Dent2014}. Their method was able to recover a face-on surface brightness profile similar to that obtained by \citet{Augereau2001}, who fitted a collisional model to scattered light observations with a parametrised radial profile for the parent belt. Despite its success, the recovered surface brightness profile was dependent on the boundaries defining the annuli in their model and the uncertainty of the fit was not well characterised. 

Furthermore, the height profile of debris disks also reveals the dynamical state of the disk. The thickness of the disk reflects the level of dynamical stirring within the disk, which can be used to constrain the presence of large embedded bodies within the disk \citep{Quillen2007, Pan2012, Daley2019}. Studies have often assumed that the scale height of a debris disk is proportional to the radius, thereby allowing the height to be parametrised with a single value known as the ``aspect ratio'' \citep{Olofsson2016, Sai2015}. However, based on scattered light observations with the Hubble Space Telescope, \citet{Graham2007} found that the vertical full width half maximum (FWHM) of the projected flux of AU~Mic first decreases with radius up to 40--50~au before it starts increasing sharply. \citet{Kalas1995} found a similar trend in Beta Pic, where the scale height appears constant up to a certain radius before it starts increasing. 

At mm wavelengths which target thermal emission, the Atacama Large Millimeter/submillimeter Array (ALMA) has also produced observations that are able to resolve the vertical structure of disks. Using ALMA observations of Beta Pic, \citet{Matra2019} were able to determine that Beta Pic's debris disk may have two dynamical components with different heights, similar to the hot and cold components of the classical Kuiper belt. Such observations suggest that the scale height of debris disks may exhibit more complex behaviour than can be characterised by a single aspect ratio, and that recovering any such variations may potentially be informative for interpreting the dynamical interactions within the disk. 

Given the importance of recovering the surface brightness and height profile in an unbiased manner, it is desirable that this information could be fitted non-parametrically. In this study, we develop a non-parametric method to recover the face-on radial surface brightness profile (brightness as a function of radius) and height profile (height as a function of radius) of debris disks that builds on the method developed by \citet{Telesco2005} and \citet{Dent2014}. We describe the algorithm in Sec.~\ref{sec:algorithm} and apply it to test cases in Sec.~\ref{sec:demonstration} to demonstrate its behaviour and performance. We then apply the algorithm to recover the surface brightness and height profile of the debris disk of AU~Mic in Sec.~\ref{sec:aumic} and compare it with results from previous studies based on parametric modelling.

\section{Algorithm}
\label{sec:algorithm}
In this section, we first describe a method to take an (noisy, PSF-convolved) image of an edge-on debris disk as input and output the surface brightness as a function of radius if the disk were to be viewed face-on (Sec.~\ref{sec:radial_method}). 
%This relies on a ``matrix inversion'' method (Sec.~\ref{sec:matrix_inversion}), which is optimised using a Monte-Carlo approach (Sec.~\ref{sec:boundaries}). 
We then describe a method to constrain the scale height of the debris disk as a function of radius (Sec.~\ref{sec:height_method}), which relies on the recovered face-on surface brightness profile and an assumption about the inclination of the disk. 
Finally, we describe a method to constrain the inclination of the disk (Sec.~\ref{sec:inc}) which could inform the range of plausible inclination assumptions used for height fitting. 

As a brief overview, the fitting method reduces an image of an edge-on disk into two 1D observables: the total vertically integrated flux as a function of distance along the midplane, and the flux in the midplane as a function of distance along the midplane (calculated by vertically integrating the flux within a small distance of $y_\text{mid}$ from the midplane). To recover the face-on surface brightness as a function of radius, the algorithm finds a radial profile that reproduces the former. Given this fitted face-on surface brightness profile, to find the height as a function of radius, the algorithm varies the height at each radial location until the right amount of flux is found in the midplane, i.e., reproducing the latter observable.

\subsection{Surface brightness profile}
\label{sec:radial_method}
The procedure described in this section aims to recover the face-on radial surface brightness profile, $f(r)$, of an edge-on, optically thin debris disk using a non-parametric approach (i.e., $f(r)$ is the surface brightness that would have been observed if the disk had been observed face-on at infinite resolution and sensitivity). Importantly, the procedure is performed under the assumption that the disk is azimuthally symmetric. 

The observed disk image is assumed to be rotated such that the projected major axis lies horizontally in the image along the $x$-axis. The general idea is to view the disk---and therefore its noiseless, PSF-convolved image, $I_\text{obs}(x, y)$---as the superposition of a set of $N$ concentric annuli with PSF-convolved images $I_{\text{ann-} i}(x, y)$, such that
\begin{equation}
    I_\text{obs}(x, y) = \sum_{i=1}^{N} I_{\text{ann-} i}(x, y),
\end{equation}
\noindent where $x$ and $y$ are the spatial coordinates in the image plane along and perpendicular to the disk's projected major axis respectively. The point $x = y = 0$ corresponds to the location of the star, and the positive $x$ direction is defined to be rightwards in the image and the positive $y$ direction upwards. 

The edges of each annulus border those of its neighbours, effectively partitioning the disk into $N$ discrete regions. Given this partitioning, it is possible to generate PSF-convolved model images of each annulus, $\bar I_{\text{ann-} i}(x, y)$, assuming that they have a uniform face-on surface brightness with normalised intensity (or ``normalised annuli''). The discretised radial profile of the disk, $\vb* F = (F_1, F_2, ..., F_{N})$, is then effectively the “weights” required of each normalised annulus:
%to reproduce the observed projected flux distribution:

\begin{equation}
    I_{\text{ann-} i}(x, y) = F_i \, \bar I_{\text{ann-} i}(x, y).
\end{equation}

It is the goal of the radial profile fitting algorithm to find $F_i$ for all annuli from $i=1$ to $i=N$ as described in Sec.~\ref{sec:matrix_inversion}.

\subsubsection{Annuli generation}
\label{sec:generate}

We generate images of annuli, $\bar I_{\text{ann-} i}(x, y)$, using a Monte Carlo approach. Each individual annulus (indexed $i$) is defined by an inner and outer boundary ($R_{i-1}$ and $R_i$) and a scale height ($H_i$). Note that scale heights quoted in this paper are standard deviations of the vertical distribution, rather than the FWHM (consistent with the convention in the literature). Points are then generated in cylindrical coordinates, $(r,\,\phi,\,z)$, of which $r$ is drawn from a ``trapezoidal distribution'' defined by
\begin{equation}
    P(r) = \begin{cases}
    \frac{2 \pi r}{\pi (R_i^2 - R_{i-1}^2)} \ , \ R_{i-1} < r \leq R_i,\\
    0 \ , \ \text{otherwise},
    \end{cases}
\end{equation}

\noindent $\phi$ is drawn from a uniform distribution between $0$ and $2\pi$ and $z$ is drawn from a normal distribution with $(\mu,\sigma) = (0,H_i)$. 
%Points generated in this way are uniform within the annuli when projected onto the $(r, \phi)$ plane. 
Points generated in this way are uniformly distributed within the boundaries of each annulus when projected onto the $z = 0$ plane. 
The coordinates of each point, $\va*{r_j}$, are then converted to Cartesian coordinates and rotated to the right inclination about the $x$ axis: 
\begin{equation}
\begin{bmatrix}
x\\
y\\
z
\end{bmatrix} = 
\begin{bmatrix}
1 & 0 & 0\\
0 & \cos{\theta} & \sin{\theta}\\
0 & -\sin{\theta} & \cos{\theta}
\end{bmatrix}
\begin{bmatrix}
r\cos{\phi}\\
r\sin{\phi}\\
z
\end{bmatrix},
\end{equation}

\noindent where $\theta$ is the inclination of the disk from face-on (i.e., $\theta = 90^\circ$ for a perfectly edge-on disk).

After rotation, the $z$-direction corresponds to the negative line-of-sight direction, the $x$-$y$ plane the sky-plane and the $x$-direction the disk’s major axis. The points are then binned into a grid of pixels on the $x$-$y$ plane using a plate scale identical to the data to be fitted to, with each point contributing equal flux, normalised such that the annulus has unit face-on radial surface brightness profile (though the annulus is viewed from an inclined angle in the image). Finally, the annulus image is convolved with the PSF (or ``beam'' at mm wavelengths) of the observation.
%, which is usually determined using an accompanying calibrator star observation. 

In practice, when the annuli simulation is to be performed a large number of times, the process can be sped up by pre-generating a large number of narrow, concentric annuli with the same inclination, height, plate scale and PSF. The image of an annulus defined by two radial boundaries is then the sum of all narrow annuli at radial locations within the boundaries. If the boundaries are chosen such that they match those of the pre-generated annuli, this method yields annuli images rapidly without compromising accuracy. This method is referred to as the ``rapid annuli'' method.

\subsubsection{Matrix inversion method for radial profile fitting}
\label{sec:matrix_inversion}
Having generated images of inclined, normalised annuli, the goal of the fitting procedure described in this section is to constrain $\vb* F$, which represents the discretised radial profile of the disk. 

The region of the image $I_\text{obs}(x, y)$ within $y_\text{max}$ away from the major axis (the line $y = 0$) can be summed along the $y$ direction to reduce the image into a one-dimensional observable, $K_\text{obs}$:
\begin{equation}
    K_\text{obs}(x, y_\text{max}) = \int_{-y_\text{max}}^{y_\text{max}} I_\text{obs}(x, y)dy,
    \label{eq:1d_flux}
\end{equation}

Taking $y_\text{max} = \infty$ (we later take a different value of $y_\text{max}$ to define a different quantity), we obtain the ``full 1D flux'', $K_\text{obs, full}(x)$, with flux as a function of $x$ only:
\begin{equation}
    K_\text{obs, full}(x) = K_\text{obs}(x, \infty).
\end{equation}

The major axis of an edge-on disk has two halves relative to the centre of the disk image. Since the algorithm assumes axisymmetry, the left and right halves of the disk image are assumed to be identical. The subsequent algorithm can therefore be applied to either half of $K_\text{obs, full}(x)$, i.e., either the $x > 0$ half or the $x < 0$ half, where $x = y = 0$ corresponds to the centre of the image, or it could be applied to the average profile of the two halves. 

%This 1D flux is the function that the radial profile fitting method is ultimately fitting to. 

A set of $N$+1 boundaries, $\vb*{X}=(X_0, X_1, \dots, X_N)$, is used to bin $K_\text{obs, full}(x)$ into $N$ regions, where $X_0$ is set to 0 (the centroid of the disk image) and $X_N$ some distance along the midplane beyond which there is negligible flux contribution. We choose the bins in $x$ for discretising the flux to be the same as the bins in $r$ which define the annuli boundaries, i.e. $\vb* X$ = $\vb* R$. This is because each annulus contributes maximally to the imaged flux at a distance equal to its radius. 

The $K_\text{obs, full}(x)$ values within each bin are averaged to yield an $N$-dimensional vector, $\vb* L_\text{obs}$, representing the ``discretised full 1D flux'': 
\begin{equation}
L_{\text{obs}}^{j} = \text{mean}(K_\text{obs, full}(x)), \ X_{j-1} < x \leq X_j,
\end{equation}

\noindent where the components of $\vb* L$ are indexed in superscripts. 

So far, our procedure has reduced the 2D input image of a debris disk into a 1D array, $K_\text{obs, full}(x)$, which is binned into a smaller 1D array, $L_{\text{obs}}^{j}$, of size $N$, which is equal to the number of annuli used to subsequently perform the fitting. 

We then also perform the same 1D projection and binning procedure to each simulated annulus image, yielding a set of discretised 1D fluxes, $\vb* L_{\text{ann-} i}$: 
\begin{equation}
    K_{\text{ann-} i}(x, y_\text{max}, h(r), \theta) = \int_{-y_\text{max}}^{y_\text{max}} \bar I_{\text{ann-} i}(x, y, h(r), \theta)dy,
\end{equation}
\begin{equation}
    K_\text{ann-$i$, full}(x, h(r), \theta) = K_{\text{ann-} i}(x, \infty, h(r), \theta),
\end{equation}
\begin{equation}
    L_{\text{ann-} i}^{j}(h(r), \theta) = \text{mean}(K_\text{ann-$i$, full}(x, h(r), \theta)), \ X_{j-1} < x \leq X_j,
    \label{eq:begin}
\end{equation}

\noindent where the dependence of the model annulus image on the disk's scale height profile, $h(r)$, and inclination, $\theta$, is explicitly denoted. 

%Observing that for $y_\text{max} = \infty$ the height and inclination of the disk have negligible impact on the full 1D flux profile, $K_\text{ann-$i$, full}$, and therefore on $\vb* L_{\text{ann-} i}$, the $h(r)$ and $\theta$ dependence disappears. 
For a given annulus, increasing its inclination relative to edge-on spreads the flux vertically away from the major axis but not horizontally, and thus does not alter its 1D flux profile, $K_\text{ann-$i$, full}$, which is integrated across $y_\text{max} = \infty$. Increasing the height of the annulus elicits a similar behaviour, whereby flux becomes more vertically spread-out in a way similar to altering the inclination. We may therefore drop the dependence of $\vb* L_{\text{ann-} i}$ on $h(r)$ and $\theta$, and compute the discretised full 1D flux of the debris disk model resulting from these annuli based only on the discrete surface brightness profile, $F_i$:
\begin{equation}
    {L_{\text{model}}^j} = \sum_{i=1}^{N} F_i \, L_{\text{ann-} i}^j,
    \label{eq:end}
\end{equation}

%\noindent where
%\begin{equation}
%L_{\text{ann-} i}^{j}(h(r), \theta) = L_{\text{ann-} i}^{j}.
%\end{equation}

In matrix form, this can be equivalently expressed as 
\begin{equation}
    \vb*{L_{\text{model}}} = M \ \vb*F,
\end{equation}

\noindent where $M_{ji} = L_{\text{ann-} i}^j$.
Each column in $M$ is then the discretised 1D flux of a (normalised) model annulus. By setting $\vb*{L_{\text{model}}} = \vb*{L_{\text{obs}}}$, we can then recover the radial profile by simply inverting $M$:
\begin{equation}
    \vb*F = M^{-1} \ \vb*{L_{\text{obs}}}.
\end{equation}

\begin{figure}
    \centering
    \includegraphics[width=8cm]{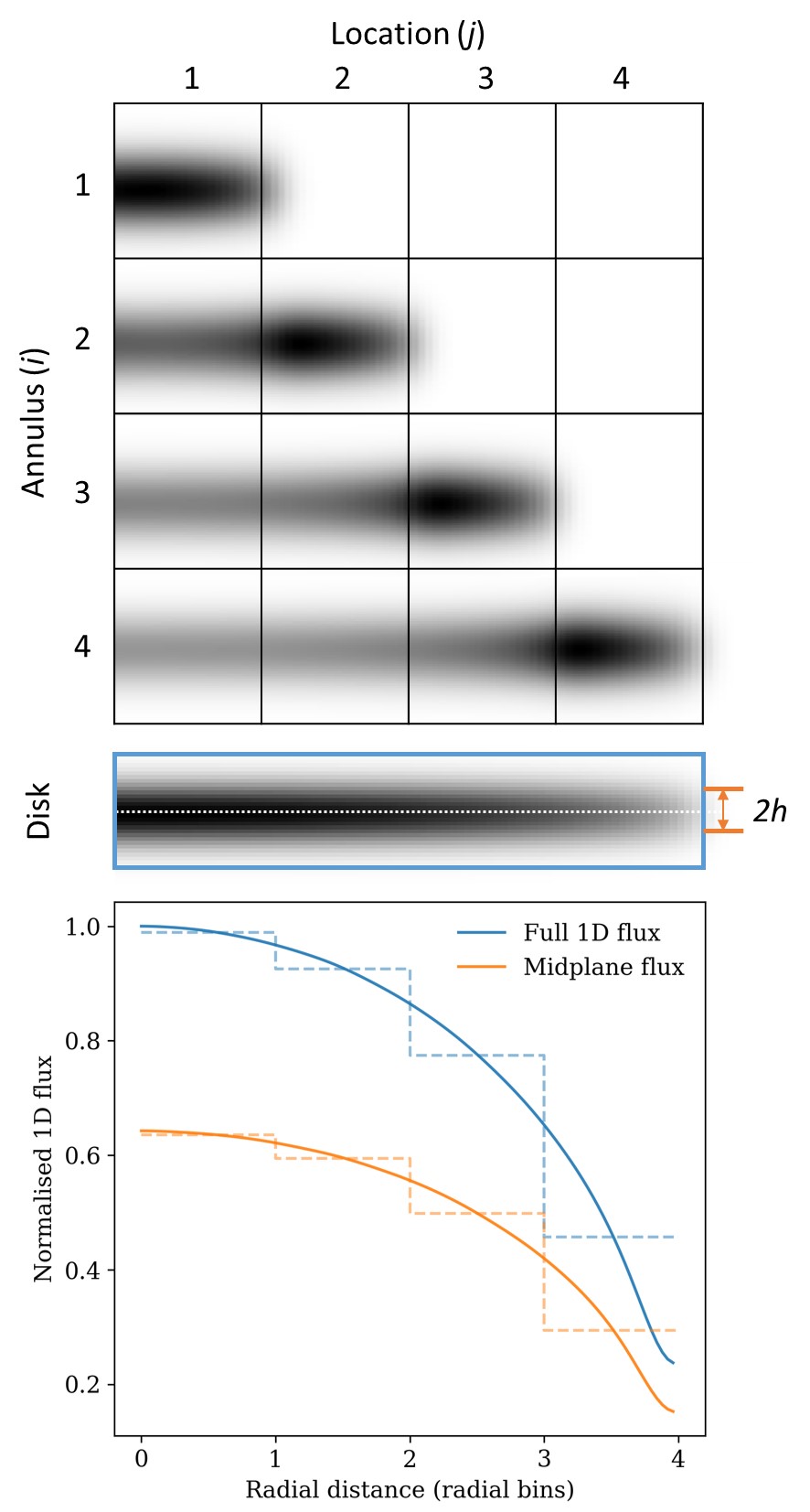}
    \caption{The top diagram shows the right half images of 4 annuli, each with a  top-hat surface brightness, whose weighted sum produces the half disk at the bottom. Images are convolved with a Gaussian PSF such that the half-disk displayed is resolved with 10 FWHMs along the major axis. The vertical lines indicate the radial boundaries of the annuli, which are the same as the boundaries used to bin their fluxes. In the disk image at the bottom, summing the flux vertically gives the ``full 1D flux'', $K_\text{obs, full}(x)$, whereas summing only the flux within distance $y_\text{mid}$ away from the midplane (dotted white line) gives the ``midplane flux'', $K_\text{obs, mid}(x)$. The bottom plot shows $K_\text{obs, full}(x)$ and $K_\text{obs, mid}(x)$ for this example. The two functions are discretised by averaging within each bin (corresponding to the cell locations in the top diagram) to give $\vb* L$ and $\vb* l$ respectively, as plotted in dashed lines. The radial profile fitting algorithm determines the weighting of each annulus needed to reproduce $\vb* L$, and the height fitting algorithm the height required to reproduce $\vb* l$. The matrix $M=M_{ji}$ is the transpose of that displayed in the top diagram. }
    \label{fig:diagram}
\end{figure}

Therefore, we can in theory recover the face-on surface brightness profile from an edge-on image by computing a matrix of the radial profiles of model-generated annuli, inverting the matrix and left-multiplying it with the observed 1D image flux. The solution obtained by such a method is unique and exactly reproduces the discretised 1D flux profile. Fig.~\ref{fig:diagram} provides a diagram of the construction of the matrix $M$ and Fig.~\ref{fig:matrix} shows an example of the matrix. 

In the radial profile fitting method used by \citet{Telesco2005} and \citet{Dent2014}, an iterative method was used to solve for $\vb* F$. Given enough iterations, the iterative method converges to the same solution to the system of linear equations as the matrix inversion method described here. In principle, the two methods are expected to achieve the same outcome. 

In this algorithm, while it may be tempting to use a large number of annuli to partition the disk so that each annulus is only one pixel wide, obtaining an almost continuous radial profile fit, subsequent testing shows that this generates unphysical artefacts (see Sec.~\ref{sec:boundaries}). Instead, we discretise the problem by using wider annuli that span a larger range in $x$. Choosing the bins in $x$ (to bin the flux in the images) to be the same as the bins in $r$ (to partition the disk into multiple annuli) is appropriate since an edge-on annulus contributes maximal flux at the projected radial location where it physically exists. Fig.~\ref{fig:diagram} illustrates the bins in $r$ in the top panel and those in $x$ in the bottom panel, which are equal. 

\subsubsection{Stellar flux}
When the star is detected in the image, the stellar flux can be included as a free parameter to be fitted to. This is easily incorporated into the fitting algorithm by including an additional annulus in the fit with a radius and height of 0, which effectively acts as a point source. Using the fitted stellar flux, the algorithm subtracts the stellar flux from the image before subsequently fitting the height of the disk. 

\subsubsection{Choice of annuli boundaries}
\label{sec:boundaries}
\begin{figure*}
    \centering
    \includegraphics[width=17cm]{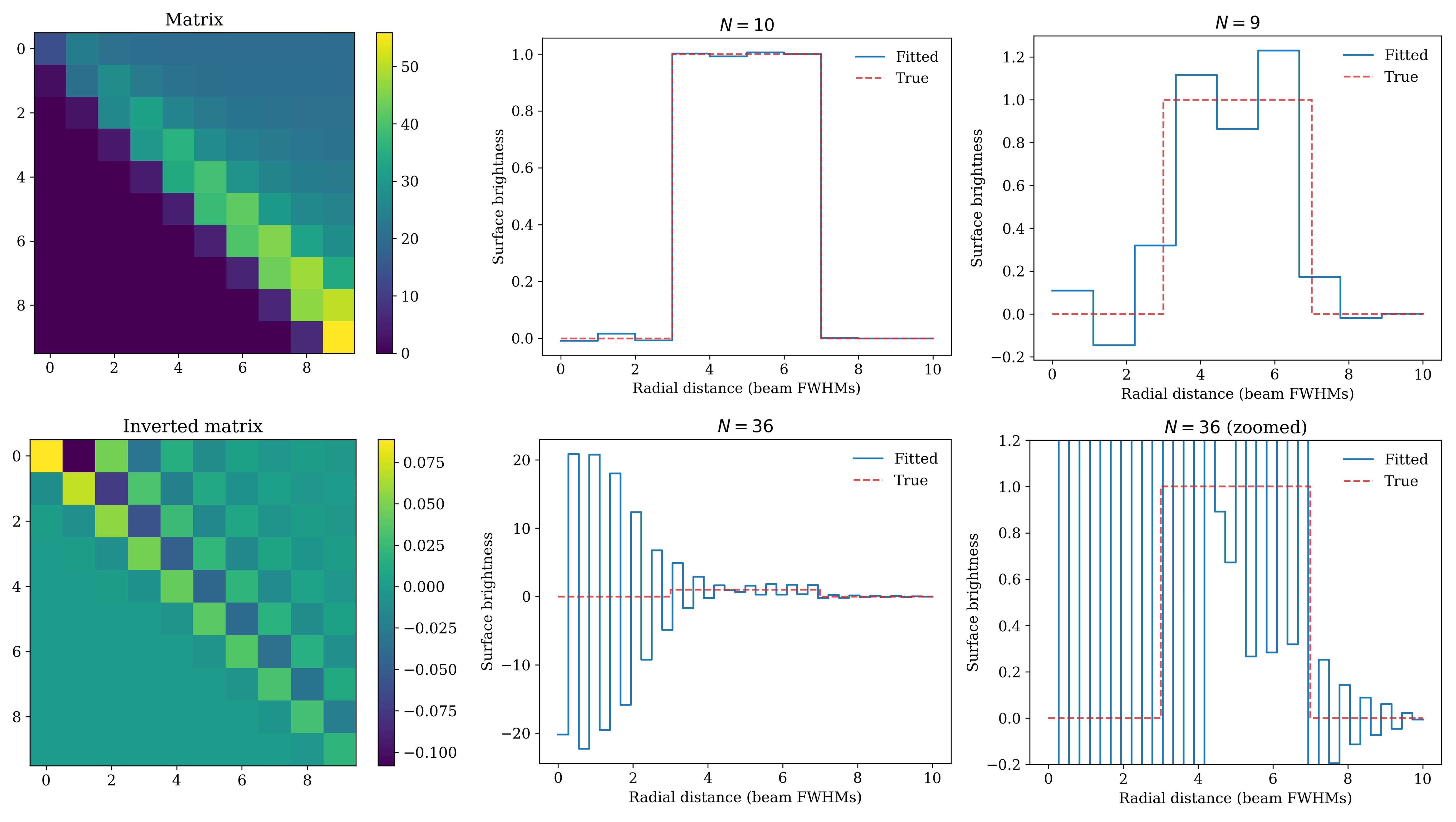}
    \caption{The two panels on the left display an example of the matrix $M$ and its inverse $M^{-1}$ for $N=10$, which are used to recover the image of a test-case disk with a top-hat surface brightness profile which contains no noise. The four panels on the right show the recovered discrete surface brightness profile, $\vb* F$, recovered with $N = 10$, 9 and 36 respectively. Although all panels are fitted to the same test case, the profiles recovered with $N = 9$ and 36 produce unphysical oscillations, whereas that recovered with $N = 10$ does not, thereby demonstrating the sensitive dependence of the fitted profile on the $N$ parameter for a radial profile with sharp edges. For extremely abrupt profiles, it is only when the annuli boundaries align with the discontinuity that the profile is accurately recovered, as demonstrated by comparing $N = $ 10 to $N = $ 9. }
    \label{fig:matrix}
\end{figure*}

The algorithm has introduced the hyperparameter $\vb* R$ that defines the boundaries of the annuli. Fig.~\ref{fig:matrix} shows examples of how changing the number of annuli defined by $\vb* R$ can alter the outcome of the matrix inversion method when fitting to a top-hat radial profile, which represents an extremely abrupt test case. This shows that the choice of $\vb* R$ is important as the fitted radial profile can be sensitively dependant on $\vb* R$ for certain underlying radial profiles. Even though all three fits exactly reproduce the observed 1D flux profile discretised to their respective number of bins, the fitted profiles are drastically different. If the model uses a large number of annuli, negative surface brightness values and unphysical oscillations may appear, as in the case of the bottom-right panel in Fig.~\ref{fig:matrix}. 
%Furthermore, each annulus is subject to a greater degree of influence from noise. 
On the other hand, if the model uses a small number of annuli, the fitted radial profile becomes overly coarse and may not have taken full advantage of the image resolution. 

Even for a reasonable number of annuli, we observe that for sharp features that are of scales smaller than the beam size, the precise location of the boundaries can still sensitively influence the solution. For a top-hat radial distribution, for example, it is only when the boundary of the annuli align exactly with the discontinuity that the top-hat feature is accurately recovered without significant artefacts. The centre-right panel in Fig.~\ref{fig:matrix} uses a similar number of annuli to the top-right panel, but does not reproduce the true profile accurately since the edges of its annuli do not align with the discontinuity. 

%We attempted to optimise $\vb* R$ by defining a metric to evaluate the plausibility of the fit by penalising negative fitted values and oscillations. The ($N$-1)-dimensional $\vb* R$ parameter space (the left- and right-most boundaries are fixed) was then explored with an MCMC to find the optimal placement of boundaries that minimise negative values and oscillations. Although such a method may converge for simple test cases such as top-hats, it cannot reasonably converge for more complicated radial profiles. Furthermore, any metric itself is likely biased as radial profile oscillations such as those due to rings in debris disks could be real. 

To achieve a less biased fitting method, we instead randomly generate many sets of $\vb* R$. $N$-1 random numbers are drawn from a uniform distribution between $R_0$ and $R_N$ and then sorted. (If the rapid annuli generation method is used, each component of $\vb* R$ is clipped to the nearest allowed boundary location.) The proposed $\vb* R$ is accepted if the width of each annulus is between a lower and upper bound to ensure that the sizes of the annuli are within a characteristic range given the number of annuli. This is set to be the range between 0.3 and 2 $R_N/N$ by default. The algorithm then fits the radial profile using each set of $\vb* R$ independently, and takes the median of all the fitted radial profiles as the best-fit solution for $f(r)$. The spread of the fits is used to estimate the boundaries within which the radial profiles should lie.

Both image noise and discontinuities in the underlying radial profile imply that there may be negative values in the fitted radial profile (see e.g. Fig.~\ref{fig:matrix}). Both image noise and discontinuities in the underlying radial profile imply that the fitted radial profile can contain negative values even though these are physically unrealistic (see e.g. Fig.~\ref{fig:matrix}). Such artefacts are mitigated by taking the best-fit profile to be the median of the profiles of many fits with randomised annuli boundaries. Since it is still possible for the median surface brightness profile to contain unphysical negative values, we set any negative values in the best-fit profile to 0. Although this may decrease the accuracy of the fit, this a necessary price to pay to obtain a physical model and to allow for subsequent height fitting. 

A number of test cases demonstrate that such a method not only converges towards a reasonable fit for most radial profiles tested within error margins, but also yields a smooth curve despite the piece-wise constant nature of each individual fit. Examples are given in Sec.~\ref{sec:examples}.

\subsection{Height distribution}
\label{sec:height_method}
The radial profile fitting algorithm reduces all data and model images into a 1D problem by projecting all fluxes onto the disk midplane. By considering the vertical flux distribution, we may also constrain the height profile of the disk. 

In addition to the discrete ``full 1D flux'', $\vb* L$, used in the radial profile fitting, we define the corresponding discrete 1D quantity called the ``midplane flux'', denoted by $\vb* l$, which is obtained by projecting only the $-y_\text{mid} < y < y_\text{mid}$ region of an image, $I(x, y)$, onto the disk's major axis:
\begin{equation}
    K_\text{obs, mid}(x) = K_\text{obs}(x, y_\text{mid}),
\end{equation}

\begin{equation}
l_{\text{obs}}^{j} = \text{mean}(K_\text{obs, mid}(x)), \ R_{j-1} < x \leq R_j.
\end{equation}

For the annuli images, however, the corresponding quantities now retain dependence on the scale height profile, $h(r)$, and inclination, $\theta$, as these two quantities determine the extent to which the flux is concentrated near the major axis:
\begin{equation}
    K_\text{ann-$i$, mid}(x, h(r), \theta) = K_\text{ann-$i$}(x, y_\text{mid}, h(r), \theta),
\end{equation}

\begin{equation}
    l_{\text{ann-} i}^{j}(h(r), \theta) = \text{mean}(K_{\text{ann-}i, \text{mid}}(x, h(r), \theta)), \ R_{j-1} < x \leq R_j.
\end{equation}

This is to be contrasted with Eqs.~\ref{eq:begin}--\ref{eq:end}, in which the full 1D flux is independent of the scale height and inclination. The midplane flux, $\vb* l_{\text{model}}^{j}$, and the deprojected face-on surface brightness profile, $\vb* F$, are then related by
\begin{equation}
    {l_{\text{model}}^j} = \sum_{i=1}^{N} F_i \, l_{\text{ann-} i}^j(H_i, \theta),
\end{equation}

\noindent where $H_i$ is the scale height of the $i$-th annulus. The scale height is assumed to be constant within each annulus. In matrix form, 
\begin{equation}
    \vb*{l_{\text{model}}} = M_\text{mid}(\vb* H ,\theta) \ \vb*F,
\end{equation}

\noindent where $M_{\text{mid}, ji}(\vb* H ,\theta) = \vb*l_{\text{ann-} i}^j(H_i, \theta)$ and $\vb* H = (H_1, H_2, \dots, H_N)$.

Unlike $M$, the new matrix, $M_\text{mid}$, is a function of both the annuli heights, $\vb* H$, and their inclination, $\theta$. The heights and inclination are covariant in the disk image and are difficult to fit simultaneously. The height fitting procedure can therefore only be carried out for an assumed inclination, and the fitting shall be repeated for a range of different inclination values within a plausible interval (see Sec.~\ref{sec:inc} for placing constraints on the inclination independent of the height fitting procedure). For an assumed inclination, we vary $\vb* H$ iteratively to converge towards the set of heights that result in $\vb*{l_{\text{model}}} = \vb*{l_{\text{obs}}}$, regenerating the matrix $M_\text{mid}$ at each iteration with the updated estimate of $\vb* H$. This iterative fitting procedure is described below. 

\subsubsection{Iterative height fitting}
\label{sec:iterative_height}
In this section, we index the annulus number with $i$, the discretised radial location $j$ and the iteration number with $k$. 
%At each height-fitting iteration $k$, the heights of each annulus $i$, denoted as $H_{i, k}$, are successively updated to get an improved estimate before moving on to the next iteration, $k+1$, where iterating over $i$ is repeated. 
As an overview of the procedure, at each iteration $k$, the height of all annuli (denoted as $H_{i, k}$) are updated to an improved estimate, starting from the outer-most annulus and moving inwards towards the inner-most annulus. The algorithm then moves onto iteration $k+1$ and repeats. 

For a given set of annuli boundaries, $\vb* R$, radial-profile fitting with the matrix-inversion method is first performed to give $f(r)$. The height-fitting algorithm is then initialised with $H_{i,0} = H_\text{initial}$ for all $i$. At each iteration, $k$, the image of each annulus $i$ is simulated with scale heights $H_{i, k}$ to give their midplane fluxes, and the discrete midplane flux of the resulting multi-annulus model is computed with 
\begin{equation}
    l_{\text{model}, k} = \sum_{i=1}^{N} F_{i} \, l_{\text{ann-} i}(H_{i, k}).
\end{equation}

Within the same iteration $k$, the algorithm then iterates over all annuli to update their scale heights. This secondary iteration begins with the outermost annulus, since each annulus $i$ contributes to $L_\text{model}^j$ and $l_\text{model}^j$ maximally at a distance equal to its physical radial location, i.e., at $j = i$, which corresponds to $R_{i-1} \leq x < R_i$. 

For an annulus $i$, the discrete midplane flux of the model at radial location $j=i$, $l_{\text{model}, k}^{j=i}$, is compared with the observed value, $l_{\text{obs}}^{j=i}$. The algorithm then calculates the new height required to compensate for a fraction, $q$, of the flux difference using a function which maps the midplane flux to the height for that particular annulus: 
\begin{equation}
    H_{i, k+1} = \mathscr{H}_i(l_{\text{ann-} i, k}^{j=i} + q(l_{\text{obs}}^{j=i} - l_{\text{model}, k}^{j=i})).
    \label{eq:replace}
\end{equation}

The height-calibration function, $\mathscr{H}_i$, in Eq.~\ref{eq:replace} is specific to annulus $i$, and is obtained by pre-generating the annulus multiple times with different scale heights, each time giving a $l_{\text{ann-} i}^{j=i}$ value. Interpolating between these points allows for a mapping between the midplane flux of an annulus at any location $j$ and the scale height. Note that $L_{\text{ann-} i}$ is unaffected when the scale height is changed as it is obtained by summing over all flux vertically for any given range in $x$. 

In Eq.~\ref{eq:replace}, the value of $q$ should be less than 1 since setting $q = 1$ would ignore the fact that other annuli $i \neq j$ may also contribute to the flux at location $j$. Empirically, $q = 0.3$ is an appropriate value to converge to the true scale height. 

In the next step, the algorithm regenerates the model $\vb* l_{\text{model}, k}$ to account for the updated height of the annulus, before moving to the inner neighbouring annulus to update its height.
%\begin{equation}
%    \vb* l_{\text{model}, k} \rightarrow \vb* l_{\text{model}, k} - F_i 
%    \vb* l_{\text{ann-} k} + F_i \vb* l_{\text{ann-} k+1}.
%\end{equation}

After all annuli heights are updated, $k$ is incremented and the process repeats. The algorithm continues incrementing $k$ until $H_{i}$ converges, giving a set of discretised scale heights $\vb* H$ for the given set of annuli boundaries. 

Similar to the case in Sec.~\ref{sec:boundaries}, the fitted height profile sensitively depends on the choice of annuli boundaries. This entire procedure is therefore repeated for a large number of randomised annuli boundaries. The median of all fits is used to estimate the best-fit height profile and the distribution of the fits is used to define the bounds within which the height profile should lie.

%\subsubsection{Rapid $\vb* l$ generation and height calibration}
%\label{sec:calibration}
%In the iterative height fitting procedure, the most computationally expensive step is the regeneration of annuli when the height is varied. The discrete midplane flux of any annulus, $\vb*l_{\text{ann-} i}(H_i)$, is a function of the annulus boundaries and height, for a given PSF, vertical summation height, $y_\text{mid}$, and inclination, $\theta$.

%When a large number of height-fitting iterations are performed for a given set of annuli boundaries, $\vb* R$, the algorithm can speed up the calculation by pre-generating a number of $\vb*l_{\text{ann-} i}(H_i)$ vectors over a range of $H_i$ values for each individual annulus $i$. Each vector forms a row in the $H_i$ -- $\vb* R$ grid, $G_i$, which is rapidly interpolated to find $\vb*l_{\text{ann-} i}(H_i)$ when $H_i$ is varied in the iterative fitting procedure. 

%With this $H_i$ -- $\vb* R$ grid comes the height-calibration functions for free. Taking the column $G_i(j=i)$, we obtain the midplane flux at $j=i$, ${l}_{\text{ann-} i}^{j=i}$, as a function of $H_i$. These points are interpolated to give the function $f_{\text{height}, i}$ used to update the height of annulus $i$ at each iteration of the fitting procedure. 

\subsubsection{Regions with low surface brightness}
The height cannot be constrained in regions where the flux is consistent with 0 within uncertainty margins, since the height is effectively undefined where there is no flux. When there exists regions with very low surface brightness, the height algorithm can optionally ignore these regions. Specifically, a model containing only these low-flux regions is first generated using the fitted surface brightness for these regions (which are near but not exactly 0) and an ``estimated height'', $h_\text{est}$. The model of the low-flux regions is then subtracted from the observation, leaving only flux from sufficiently bright regions to obtain a robust height fit. 

The extent of the low-flux region requires user input. Empirically, we recommend setting this to any region where the fitted flux is consistent with 0 within 3$\sigma$. Similarly, the estimated height for these low flux regions also requires manual input. This estimate need not be accurate as their flux contribution is relatively low, and it suffices to use approximations such as the FWHM of a vertical line profile minus the FWHM of the beam size. 

%Including these regions in the fits would produce unphysical estimates with large uncertainties that would bias the fitted height in regions with non-zero flux due to the discrete nature of the method (an annulus in a fit may span both zero flux and non-zero flux regions). 

\subsection{Constraints on inclination}
\label{sec:inc}
\begin{figure}
    \centering
    \includegraphics[width=9cm]{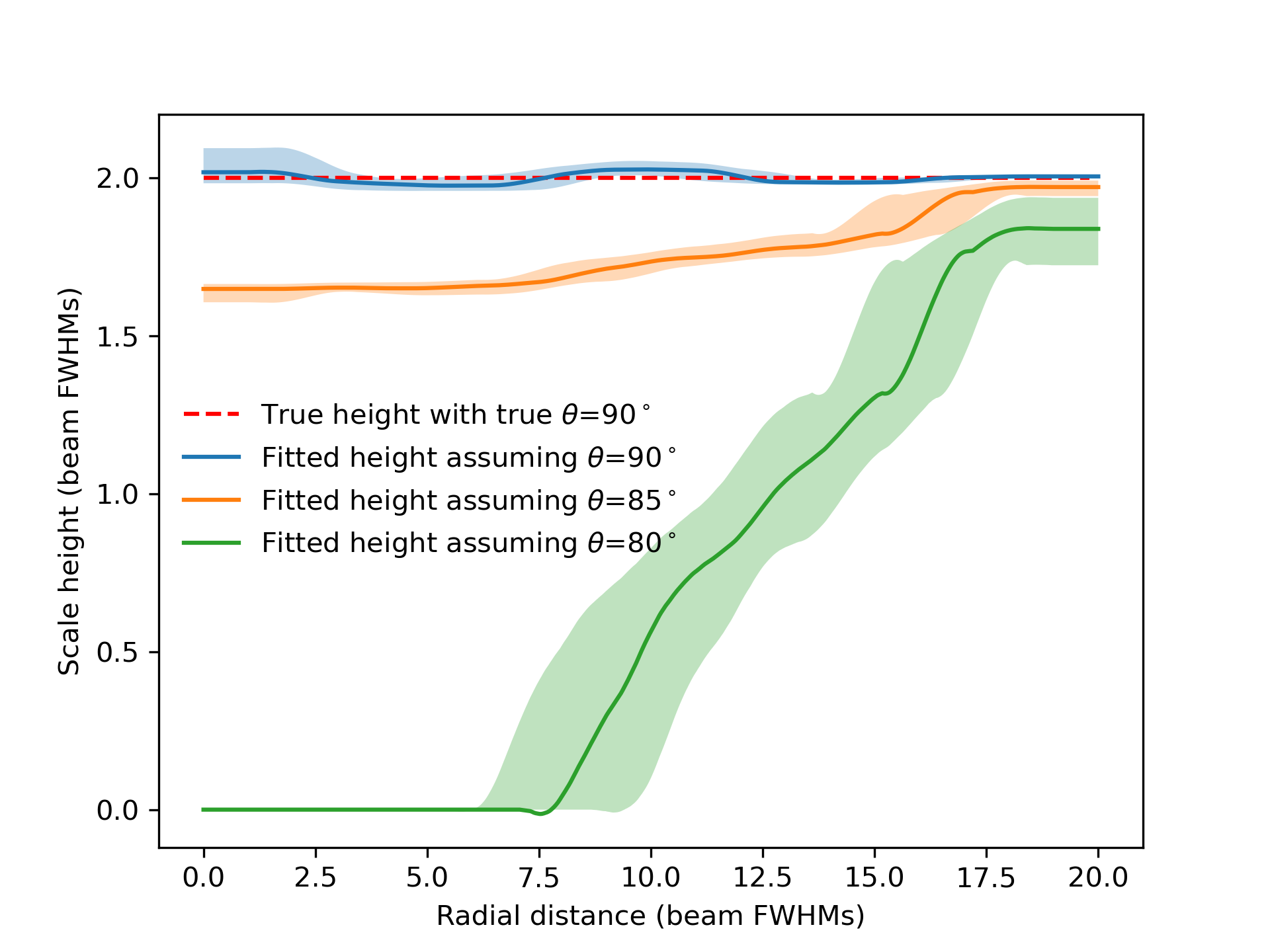}
    \caption{This plot illustrates the degeneracy between inclination and scale height. It concerns a simple test case with uniform face-on surface brightness, uniform scale height, infinite S/N per beam and viewed perfectly edge-on. The recovered height profile fitted with $N=5$ decreases as the assumed inclination decreases. }
    \label{fig:h_inc}
\end{figure}

%The degree to which a disk's flux is concentrated around the midplane sets an upper bound on $\lvert \theta - 90^\circ \rvert$. 
Unlike surface brightness profile fitting, there is no unique solution to the height profile due to its degeneracy with inclination. Fig.~\ref{fig:h_inc} provides an example of how different height profiles are recovered under different inclination assumptions when fitting. In particular, as the assumed inclination decreases (i.e. further away from edge-on), the recovered height profile decreases. This reflects the fact that lower inclinations disperse flux further away from the major axis of the disk, therefore requiring a flatter disk to compensate for the midplane flux. When the assumed inclination is sufficiently low, the height profile required to reproduce the observed midplane flux becomes 0 at some radial locations, making the disk ``flat''. At even lower inclination, there becomes no height profile that can reproduce the observed midplane flux. Such inclinations are therefore incompatible with the observation, thereby setting a constraint on the inclination. 

While the inclination constraint obtained in such a way is informative, in practice it may also be useful to have a method that can set inclination constraints independent of the height fitting method. Such an independent inclination fitting procedure can be performed before subsequent height fitting to inform the range of plausible inclination assumptions. 

Since the flux of a disk is maximally concentrated in the midplane of the image (i.e. within the region $-h_\text{mid} < y < h_\text{mid}$) when the disk is perfectly flat (i.e., $\vb* H = \vb* 0$) and the inclination $\theta$ = $90^\circ$, an allowed value of $\theta$ must then satisfy
\begin{equation}
\label{eq:theta}
    l_{\text{model}}^{j}(\theta, \vb*H = \vb*0) > l_{\text{obs}}^{j}
\end{equation}
at all locations, $j$. 

For optically thin disks, only $\lvert \theta - 90^\circ \rvert$ is constrained, e.g., an inclination of $85^\circ$ is indistinguishable from $95^\circ$. 
For simplicity, we only label the inclination with values below $90^\circ$ throughout this paper. 

Given a fitted radial profile, $f(r)$, a model image of a flat disk (i.e. height equal to 0 everywhere) can be generated for any desired inclination, $\theta$. To obtain a lower bound for $\theta$, the algorithm starts from $\theta = 90^\circ$ and iteratively decreases $\theta$ until finding the lowest $\theta$ at the desired precision that satisfies Eq.~\ref{eq:theta}. 
%At each iteration, $\vb*l_{\text{obs}}(\theta, \vb*H = \vb*0)$ is compared with $\vb* l_{\text{data}}$, until the lower bound for $\theta$ is constrained to the required accuracy. 

We point out that the ``midplane region'' (characterised by $y_\text{mid}$) which defines the midplane flux must be sufficiently large such that the central hole of an inclined disk (that is not perfectly edge-on) is fully encompassed by the midplane region, i.e. $y_\text{max} \geq r_\text{max} \sin{\theta}$, where $r_\text{max}$ is the radial location beyond which the surface brightness is negligible. This condition is necessary in order for the midplane flux to decrease as the height increases, an assumption on which the inclination fitting method relies. 

\subsection{Applicability}
The requirements on the disk in order for both the surface brightness and height fitting algorithms to apply is that the disk must be optically thin and azimuthally symmetric. For the height fitting algorithm to apply, it is additionally required that the vertical distribution of material can be approximated by a Gaussian. While the algorithm is primarily designed to analyse debris disks, other disks that satisfy the same requirements imaged at optically thin wavelengths may also be analysed with this method. 

The requirement on the wavelength of observation used as input is that it corresponds to the distribution of thermal emission. The most common thermal emission observations of debris disks are in the mid- and far-IR and mm wavelengths. Scattered light observations are often subject to non-isotropic scattering effects which may require additional modelling to account for when trying to recover the true distribution of underlying material. However, the method may still be directly applicable under the assumption that the scattering is isotropic. The algorithm also performs fitting in image space, and any visibility data must first be converted into image-space data before applying the algorithm. 

The surface brightness fitting method does not depend on knowledge about the inclination and is therefore valid for deconvolving images at any inclination, including images of face-on disks. 
%However, alternative methods that directly deconvolve the face-on, PSF-convolved flux profile are likely more useful.
The height profile method requires knowledge of the vertical distribution of emission, which becomes decreasingly accessible as the inclination decreases. This method is therefore most applicable to edge-on or nearly edge-on disks (i.e., relatively high inclinations of $\gtrsim 75^\circ$).

\subsection{Implementation}
\texttt{Rave} (Radial And Vertical Edge-on disk fitter) is an object-oriented implementation of the surface brightness and height profile fitting algorithm in the \texttt{Python} language. \texttt{Rave} and associated demonstrations are available at \href{https://github.com/yinuohan/Rave}{github.com/yinuohan/Rave}.

\section{Demonstration}
\label{sec:demonstration}
\subsection{Test cases}
\label{sec:examples}
This section describes the accuracy and precision of the recovered radial profile, height distribution and inclination constraints by exploring test cases with known disk parameters. 

We simulated images of edge-on disk observations by first defining a radial profile, $f_\text{true}(r)$, and height distribution, $h_\text{true}(r)$, which are used to generate simulated images using the Monte Carlo approach described in Sec.~\ref{sec:generate}. We convolved the images with Gaussian PSFs and added normally distributed noise which is convolved with the beam. Length scales in these test cases are defined using multiples of the Gaussian beam's FWHM.
%(a source with twice the angular size observed with twice the beam size gives the same image). 

For each simulated image, the radial profile fit was performed 100 times with randomised annuli boundaries, under the condition that all annuli widths must be between 0.3 and 2 times the mean width, $r_\text{max}/N$. The median of all fits was used as the best-fit model.

\subsubsection{Estimating uncertainties}
%In the radial profile fitting method, the algorithm produces an exact match to the data as represented by the discrete 1D flux, $\vb* L$. This implies that the $\chi^2$ of the data versus the model is 0. However, since the data contain noise, the recovered surface brightness must therefore be overfitted. To mitigate the overfitting issue, we first obtain the best-fit model that exactly reproduces the data and then add noise to the model-generated image. 
%Actually, I realised this can't really be mitigated. All we can do is to increase the uncertainty region hoping that it contains the true distribution. 

There exists two contributions towards the uncertainty of the radial profile and height fitting algorithms. Firstly, observations contain noise. This is characterised by the signal-to-noise ratio (S/N) per beam as defined in Sec.~\ref{sec:s/n}.

Secondly, even when fitting to a noiseless observation, the algorithm exhibits an intrinsic uncertainty. This is evident from the fact that a change in annuli boundaries used to perform the fit results in a change in the recovered profile. 
Estimating this second contribution (i.e., intrinsic scatter) is relatively simple, as we can use the 16\% and 84\% quantiles among all the fits performed to define the effective 1$\sigma$ uncertainty. To also incorporate the first contribution (i.e., image noise), we need to introduce simulated noise and repeat the fit on these simulated observations. 

Specifically, we use the best-fit model to generate a model image of the disk which does not contain noise. We then perform an additional 100 fits, adding Gaussian noise to this model image before performing each fit. The Gaussian noise per beam added is equal in magnitude to the background noise in the observation. We use the 16\% and 84\% quantiles among this second set of 100 fits is used to estimate the 1$\sigma$ uncertainty accounting for both contributions. We refer to the first set of 100 fits to the observation as the ``first fit'', and this second set of 100 fits to the model as the ``second fit''. 

The uncertainty region defined in this way implies that at each radial location, 32\% of models predict a value that lies outside this 1$\sigma$ range despite having an exact fit to the observed profile (discretised by their respective annuli boundaries). The reason that the true distribution likely lies inside the range is that many of these fits will have sharp discontinuities, whereas true distributions have radial profiles in which neighbouring annuli are correlated, implying that the profile is somewhat smooth. 
In the following sections, we show that empirically the 1$\sigma$ range defined in this way provides a good measure of the range of possible model surface brightness profiles that can explain a given observed profile. We thus refer to this as the range of possible models.

\subsubsection{Defining S/N}
\label{sec:s/n}
A natural way to characterise noise in this context is the ``signal-to-noise (S/N) per beam'', which is defined as the sum of the flux within the area of a beam divided by the root-mean-square (RMS) noise per beam. 

Another useful parameter is the mean S/N per beam across the disk, which is defined by the mean flux per beam divided by the RMS noise per beam. To compute the mean flux per beam, we must define a region containing the disk's flux and count the number of beams required to cover the disk. This region is defined as any location that is within $y_\text{max}$ away from the disk's major axis and $x_\text{max}$ away from the line $x = 0$ (i.e., vertical line passing through the central star). Since it is not the primary concern of this study to determine the outline of the disk, we leave it up to the user to (either manually or programmatically) specify $y_\text{max}$ and $x_\text{max}$ that excludes the regions in which there is negligible flux contribution beyond noise.

\subsubsection{Noise correlation}
\texttt{Rave} performs fitting in image space, and so visibility data (e.g., ALMA mm observations) must first be converted to image-space images before applying \texttt{Rave}. Observations taken in image space and visibility space carry different noise structures: noise is often largely independent between pixels at wavelengths observed directly in image space (e.g., mid-IR observations), whereas it is usually correlated on the scale of the beam size in images reconstructed from visibility data (e.g., mm observations). In the test cases presented here, we simulated noise analogous to mm observations. In practice, simulating noise with independent pixels and achieving the same effective S/N per beam does not significantly alter the outcome of the fitting. This is because the algorithm discretises the flux into radial bins (i.e., annuli) that are typically larger than the beam size, thereby discounting the effect of correlations on scales smaller than the beam size.

\subsubsection{``Gaussian'', ``smooth'' and ``abrupt'' test cases}
\label{sec:test_cases}
In the following sections, we apply \texttt{Rave} to three test cases which we refer to as the ``Gaussian'', ``smooth'' and ``abrupt'' test cases respectively. Their surface brightness profiles are given by 

\begin{equation}
    f_\text{gaussian}(r) = g(r, 10, 4),
\end{equation}
\begin{equation}
    f_\text{smooth}(r) = 4g(r, 0, 6) + 2g(r, 10, 4) + g(r, 14, 10),
\end{equation}
\begin{equation}
    f_\text{abrupt}(r) = \begin{cases}
    2 \ , \ 8 \leq r < 12,\\
    1 \ , \ 4 \leq r < 8 \ \text{or} \ 12 \leq r < 16,\\
    0 \ , \ \text{otherwise},
    \end{cases}
\end{equation}

\noindent where
\begin{equation}
    g(r, \mu, \sigma) = \exp(-\frac{r-\mu}{\sigma^2}),
\end{equation}

\noindent and their height profiles are given by

\begin{equation}
    h_\text{gaussian}(r) = 2g(r, 10, 4) + 1,
\end{equation}
\begin{equation}
    h_\text{smooth}(r) = g(r, 0.2, 5) + 6g(r, 8, 5) + 10g(r, 20, 10),
\end{equation}
\begin{equation}
    h_\text{abrupt}(r) = \begin{cases}
    3 \ , \ 8 \leq r < 12,\\
    2 \ , \ 4 \leq r < 8 \ \text{or} \ 12 \leq r < 16,\\
    \text{undefined (surface brightness is 0)}, \text{otherwise}.
    \end{cases}
\end{equation}

\noindent The units of $h$ and $r$ are both the FWHM of the PSF. 
%The face-on surface brightness, $f$, is in arbitrary units. 
The face-on surface brightness, $f$, is in units of S/N per beam. In the examples in the following sections, we multiply $f_\text{gaussian}(r)$, $f_\text{smooth}(r)$ and $f_\text{abrupt}(r)$ by a normalisation factor of 10.8, 2.1 and 4.8 respectively such that all test cases have a mean S/N per beam of 10 across the disk, with $y_\text{max}$ defined as 6 times the beam's FWHM and $x_\text{max}$ 20 times that. All disks have an inclination of $90^\circ$.

\subsection{Performance}
\subsubsection{Surface brightness}
\label{sec:radial}

\begin{figure*}
    \centering
    \includegraphics[width=18cm]{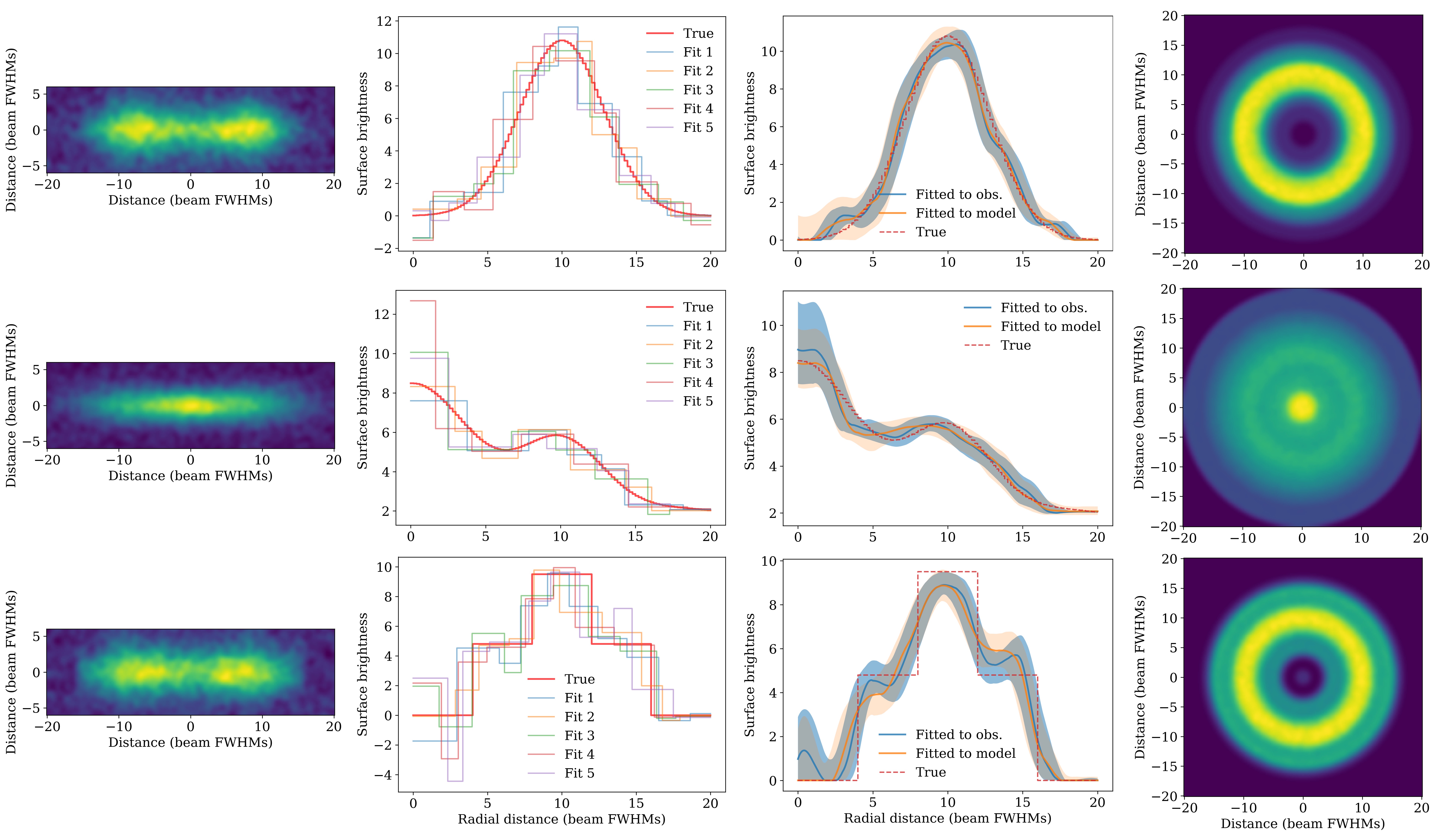}
    \caption{This figure illustrates the radial profile recovery method for the Gaussian (row~1), smooth (row~2) and abrupt (row~3) test cases. The input images (column~1) are generated with known radial profiles, which are over-plotted in columns~2 and 3 to evaluate the performance of the method. Five example individual fits (column~2) performed with randomised annuli boundaries are plotted to illustrate the fitting procedure. Column~3 shows the median radial profile and range of possible models. The resulting face-on model images convolved with the PSF are shown in Column~4. The Gaussian and abrupt test cases are fitted with $N = 10$ and the smooth test case with $n = 7$. }
    \label{fig:radial}
\end{figure*}

Fig.~\ref{fig:radial} illustrates the fitting procedure applied to the three test cases from Sec.~\ref{sec:test_cases}. The individual fits (Fig.~\ref{fig:radial} column~2) largely trace the true distribution but exhibit some degree of scatter, reflecting the dependence of the recovered radial profile on the choice of annuli boundaries. However, with 100 iterations of independent, randomly selected boundaries, the median of all individual fits is able to accurately reproduce the true distribution of the Gaussian radial profile (Fig.~\ref{fig:radial}, column~3). Even with the inclusion of more sophisticated substructures in the smooth radial profile test case, the algorithm is still able to converge to a reasonably accurate solution. 

A quantity of scientific importance for debris disks is the sharpness of the inner and outer edges of the radial profile, which is often closely linked to the level of dynamical stirring in the disk. We tested the ability of the algorithm to resolve very sharp features with the abrupt test case as an extreme example. While the algorithm did not reproduce the precise step-like nature of the radial profile variations, the fitted radial profile still recognised these substructures, albeit more smoothed out, and the true distribution is largely contained within the range of possible models. This kind of smoothing behaviour is unavoidable for features smaller than the beam size.%, and therefore the fitted profile may be seen as a lower bound on how sharp a disk's edges are in reality. 

In Fig.~\ref{fig:radial} column~3, we observe that the difference is small between the profile fitted to the original observation (``first fit'') and that fitted to the model generated from the first fit (``second fit''), even though the range of possible models of the second fit incorporates effects of independent random noise. Indeed, testing shows that uncertainty due to intrinsic scatter of the method usually dominates. In practice, it is nonetheless useful to perform both fits to quantify the effects of both sources of uncertainty. 

Across the test cases, uncertainties tend to be larger near the inner regions of the disk (this may be less obvious once negative values are truncated as described in Sec.~\ref{sec:boundaries}). This behaviour is not surprising, since the outermost regions receive flux contributions from material only at the largest radial distance from the star, whereas flux from material at all distances contribute to the innermost regions. The fitted radial profile therefore reflects the propagation of errors from the outer to inner regions along the disk midplane.

\subsubsection{Interpreting the range of possible models}
\label{sec:uncertainty}
The range of possible models in our fits carry different meanings from the uncertainty regions of parametric fitting. Rather than being the uncertainty on one particular model as is commonly the case for parametric fitting, each shaded region in Fig.~\ref{fig:radial} indicates a range that is expected to contain the true profile regardless of its shape (i.e., functional form). 

More specifically, if the true profile is smooth (e.g., the Gaussian test case), then the true profile is expected to be close to the ``median model'' as indicated using solid lines in the fits in Fig.~\ref{fig:radial}. If the true profile is sharp (e.g., the abrupt test case), then it would deviate significantly from the median model, however it should still be largely contained within the shaded region, as demonstrated in Fig.~\ref{fig:radial} row~3, column~3. 

In reality, there is no simple way of inferring whether the true profile is smooth or sharp. In order to have a shaded region that takes into account the full range of possibilities of sharpness of the true profile, we need to invoke the relatively large range of possible models. The range of possible models therefore does not imply that a profile that takes on the 1~$\sigma$ upper bound everywhere, for example, can reasonably reproduce the observation. Instead, it means that the surface brightness at a given radial location could be that high, but if so this would require a correspondingly lower point nearby.

\begin{figure}
    \centering
    \includegraphics[width=8cm]{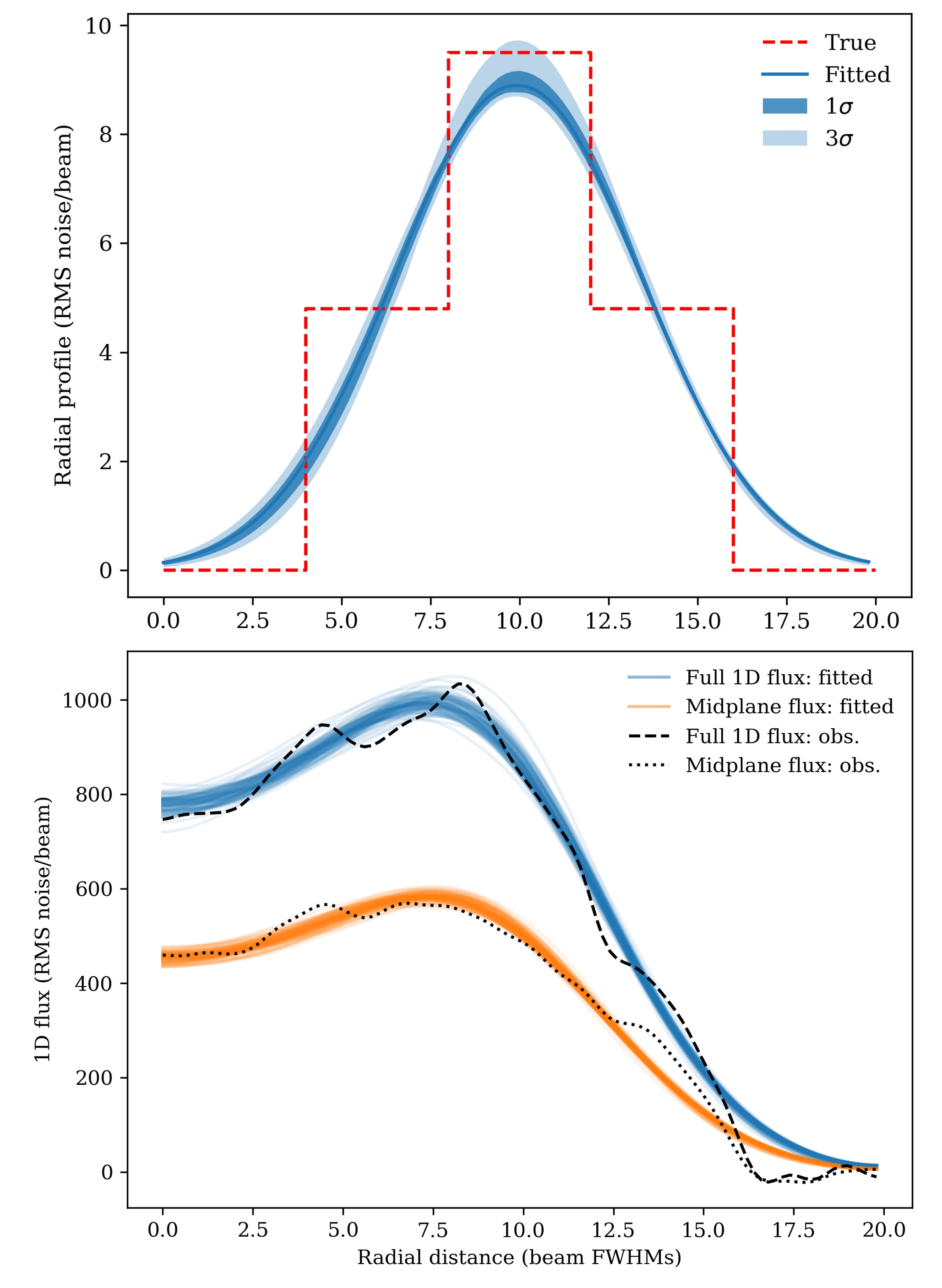}
    \caption{This figure illustrates that fitting a radial profile parametrically can significantly underestimate the uncertainties. The top panel shows the outcome of fitting to the abrupt test case but assuming that the radial profile is Gaussian. Even at the 3$\sigma$ level, the uncertainty margins do not contain the true distribution. The bottom plot shows two 1D observables from 100 models drawn from the posterior distribution, which are compared to those of the simulated observation being fitted to. }
    \label{fig:parametric}
\end{figure}

To contrast the range of possible models with the uncertainties from parametric approaches, we used an MCMC method to parametrise and fit to the abrupt test case. Fig.~\ref{fig:parametric} shows the outcome of the fit when parametrising the radial profile as a Gaussian and using the sum of the $\chi^2$ of the full 1D flux and midplane flux as the metric. Although the two 1D observables are reasonably reproduced, the uncertainties of the fitted profile are significantly underestimated. Indeed, these uncertainties refer to the those under the assumption that the assumed radial profile is true, which is not the case here. The range of possible models, on the other hand, provide a more realistic estimate, largely containing the true distribution even for the abrupt test case. 

%These test cases therefore demonstrate that for a range of features in the radial profile, and with a reasonable S/N, the algorithm is able to recover the radial profile to a reasonable degree of accuracy given an image of an edge-on disk. 

\subsubsection{Inclination constraints}
%The previous height fitting algorithm relies on assumptions of the disk's inclination. Without additional sources of constraints on the inclination, we are also able to place a lower bound on``how edge-on'' the disk may be from the vertically unresolved image. 

To demonstrate the behaviour of the independent method to constrain inclination, we applied the method to the smooth test case described in Sec.~\ref{sec:test_cases}, except with a range of different near edge-on inclinations. 

\begin{figure}
    \centering
    \includegraphics[width=7cm]{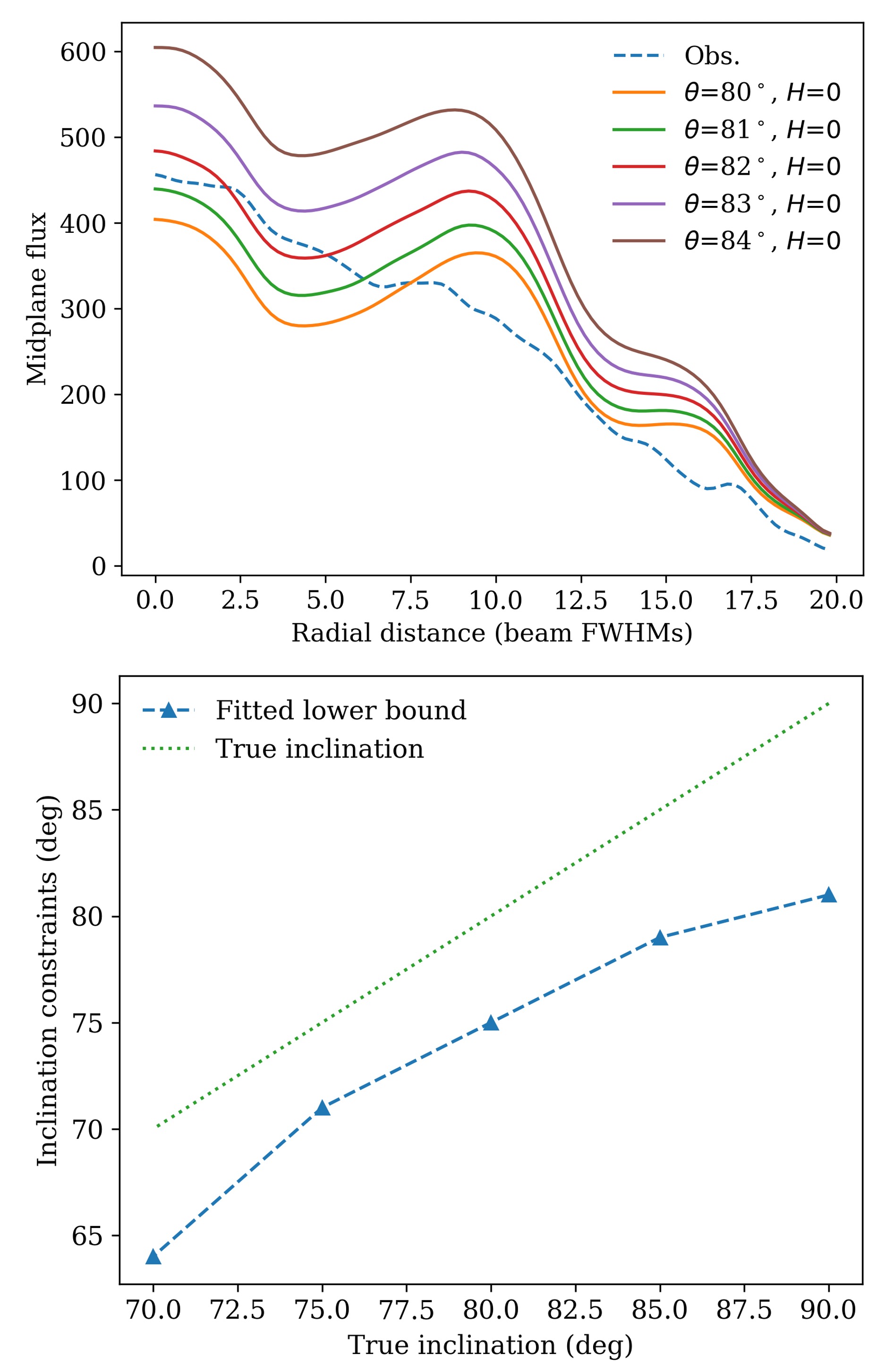}
    \caption{This figure demonstrates how a lower bound on the inclination may be obtained before fitting the scale height. 
    In the top panel, the dashed blue line represents the midplane flux of the smooth test case. Over-plotted are a set of midplane fluxes simulated using the radial profile fitted from the smooth test case, but with $H = 0$ and at various inclinations. The midplane flux must be greater at every location for a flat disk than what is observed, and so the lower bound for the inclination is determined as $82^\circ$. 
    The bottom panel summarises the lower bounds obtained in this way for observations simulated over a range of inclinations. Each data point is obtained using a plot similar to the one in the top panel. }
    \label{fig:inclination}
\end{figure}

The top panel of Fig.~\ref{fig:inclination} shows the midplane flux of flat disk models generated at several different inclinations with the radial profile fitted non-parametrically to the smooth test case. The noisy blue line is the midplane flux of the smooth test case (Sec.~\ref{sec:test_cases}) at $90^\circ$ inclination. Comparison between the lines shows that at any inclination below approximately $82^\circ$, even a flat disk cannot produce a sufficiently high midplane flux at every location to reproduce the simulated observation within uncertainties, therefore the lower bound for the inclination was determined to be approximately $82^\circ$. 

We repeated this exercise over a range of different inclinations for the smooth test case and display the constraints obtained in the bottom panel of Fig.~\ref{fig:inclination}. The lower bounds placed across all inclinations tested are valid and differ from the true inclination by only a few degrees for this height profile, $h_\text{smooth}$, providing a basis for choosing the range of inclination assumptions to use for subsequent height fitting in the absence of other inclination constraints available. 

%Since the disk flux is maximally concentrated near the midplane for an infinitesimally thin disk, it is expected that the midplane flux of such a model should always be greater than the observed midplane flux at all radial locations, unless when the disk inclination is so low (i.e. far from edge-on) that even a completely flat disk cannot concentrate enough flux near the midplane. Therefore, if any model midplane flux crosses the observed midplane flux, the inclination of that model must be too low, thereby setting a lower bound on the inclination. 

\subsubsection{Height distribution}
\label{sec:height}

\begin{figure*}
    \centering
    \includegraphics[width=18cm]{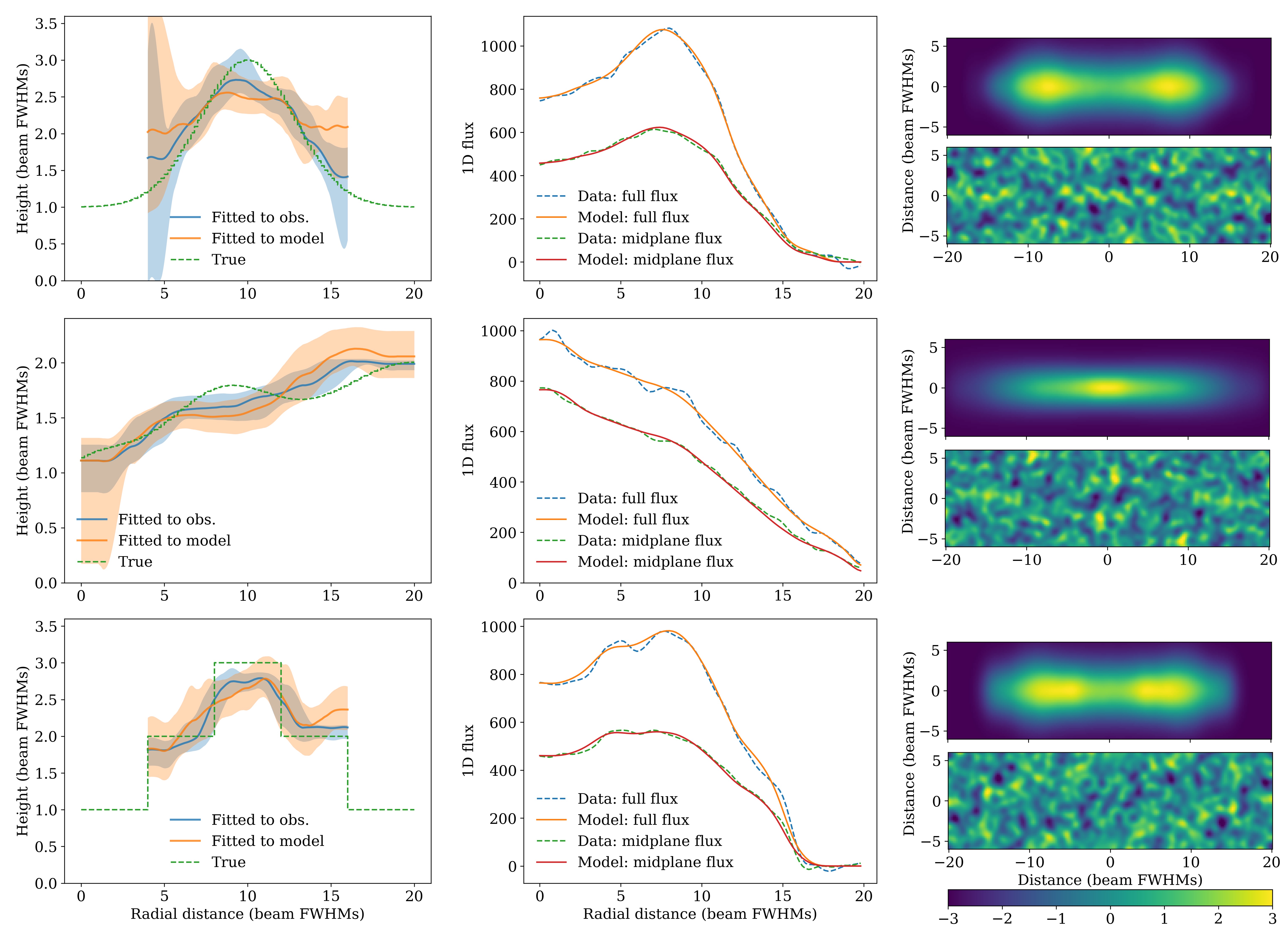}
    \caption{This figure demonstrates height fitting using the three test cases presented in Fig.~\ref{fig:radial}. Rows~1 to 3 correspond to the Gaussian, smooth and abrupt test cases respectively. Column~1: The best-fit height profile with the range of possible models fitted with $N = 7$ for the Gaussian and abrupt test cases and $N = 5$ for the smooth test case and assuming $\theta = 90^\circ$. The height for the low-flux regions are set to one FWHM of the PSF. Column~2: The image summed onto the midplane ($K_\text{obs, full}(x)$), and the flux within 2 beam FWHMs away from the midplane summed onto the midplane ($K_\text{obs, mid}(x)$), for both the input image and fitted model. $y_\text{mid}$ is defined as 2 FWHM of the beam. Column~3: Edge-on view of the model using the fitted radial profile and height distribution. Column~4: Residual of the input image after subtracting the best-fit model. The colourbar is in units of standard deviations of the background noise. }
    \label{fig:height}
\end{figure*}

To evaluate the performance of the height fitting method, we fitted height profiles for the same test cases as in Sec.~\ref{sec:radial}. The results of the height profile fit assuming $90^\circ$ inclination (identical to the true inclination) are presented in Fig.~\ref{fig:height}. Height fitting relies on knowledge of the radial profile and inclination, and any errors in the fitted radial profile displayed in Fig.~\ref{fig:radial} would therefore be propagated to the heights. 

Overall, the recovered height profiles provide a reasonable estimate of each disk's scale height, and is able to recognise broad variations in the scale height over radial distance from the star. However, finer details are more difficult to reproduce. For example, small-amplitude oscillations in the smooth case and very sharp features in the abrupt case are not accounted for in the best-fit heights. 

Previously, it was possible to generate the face-on view of the disk without knowledge of the height. With both the fitted radial profile and height distribution, it is possible to generate an edge-on model image (Fig.~\ref{fig:height} column~3). Subtraction from the noisy input image confirms that the models accurately fit the data, with no significant regions of residuals remaining. 

Column~2 in Fig.~\ref{fig:height} provides additional insight into the fitting procedure and its challenges. Both of the key observables, the full 1D flux, $K_\text{obs, full}(x)$ (as plotted in Fig.~\ref{fig:height} column~2, which is discretised to give $\vb* L_\text{obs}$) and the midplane flux, $K_\text{obs, mid}(x)$ (as plotted in Fig.~\ref{fig:height} column~2, which is discretised to give $\vb* l_\text{obs}$), are reproduced in the best-fit model for each test case. 

Indeed, finer features with small amplitudes or at small spatial scales only contribute to minor variations in the image and thus the two observables. Their effects are diluted by flux contributions from a range of radial locations, all with different heights, which is further masked by convolution with a PSF kernel and noise. For example, the Gaussian and abrupt test cases share the same broad radial variations, but differ in the abruptness of the variations. Their resulting images and 1D fluxes, however, are quite similar. These detailed features in surface brightness and height profiles are therefore encoded as very fine variations in the observed 1D fluxes, contributing to the challenges faced by the fitting method. 

% We also point out that the above fitting procedure assumes that the disk is vertically isothermal, in which case the vertical height of the emission directly reflects the vertical height of flux-emitting material. 

We conclude that the height fitting algorithm is able to recover the overall scale height of the disk and any broad radial variations, but may not trace as fine features as does the radial profile fitting algorithm.

\subsubsection{Choice of number of annuli, $N$}
While randomising annuli boundaries as discussed in Sec.~
\ref{sec:boundaries} mitigates the dependence of the fit on annuli location, the number of annuli used to perform the fit remains an important hyperparameter in the radial profile and scale height fitting algorithm. The choice of $N$ presents a classic problem reminiscent of the bias-variance trade-off in the context of machine learning \citep{Briscoe2011}. Similar to the case of fitting only one iteration (Sec.~
\ref{sec:boundaries}), partitioning the disk too finely introduces too many degrees of freedom, fitting to the noise to produce an unphysical model, while having too few annuli limits the resolution of the recovered radial profile, smoothing out key features that may be important for the science. 

\begin{figure*}
    \centering
    \includegraphics[width=18cm]{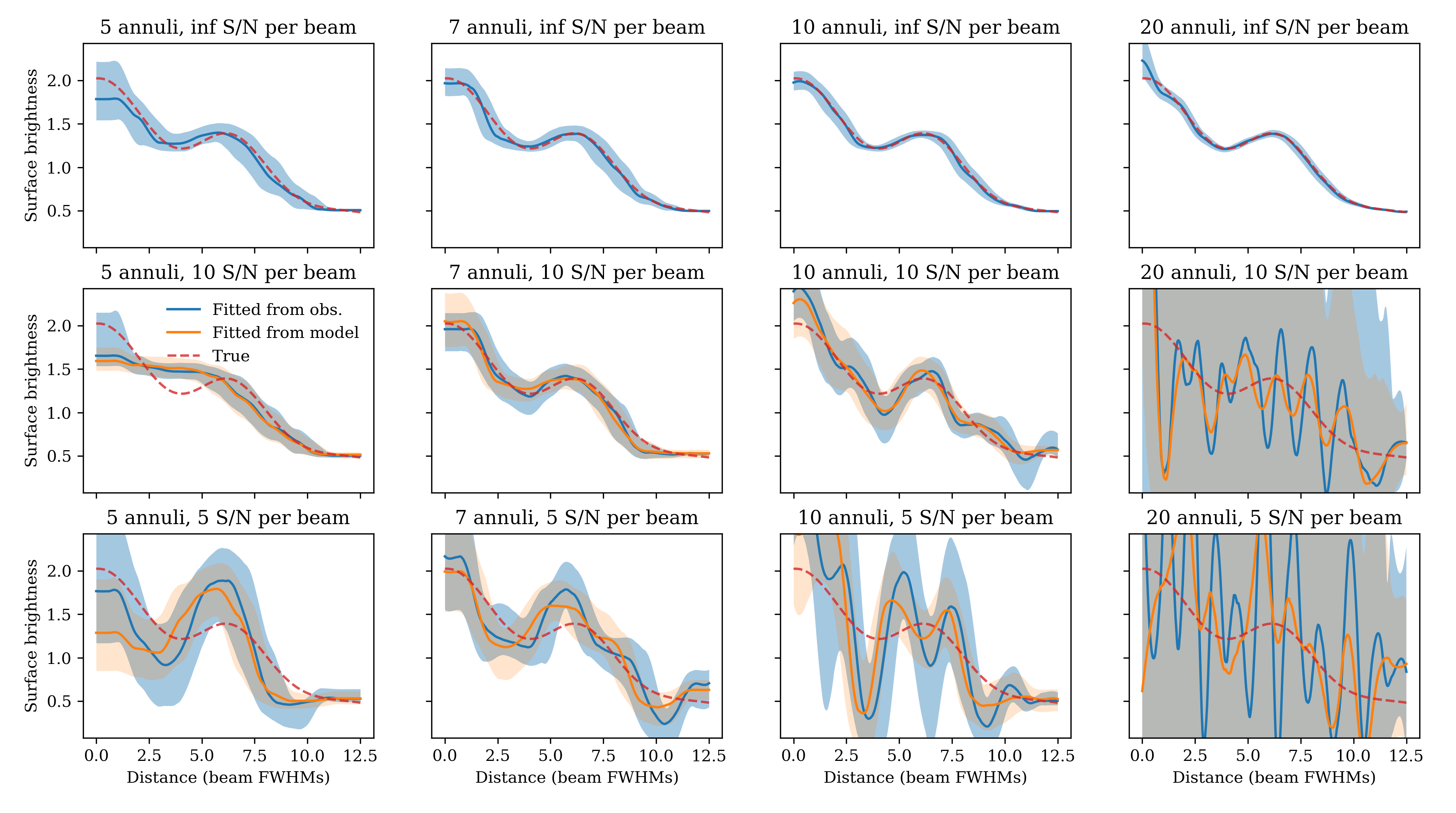}
    \caption{The set of plots in this figure demonstrate the trade-off between resolution and robustness against noise when choosing $N$. Rows 1--3 attempt to recover the face-on surface brightness profile of the smooth test case with S/N per beam equal to infinity (no noise), 10 and 5 respectively. Each column uses a different $N$. }
    \label{fig:n_annuli}
\end{figure*}

Fig.~\ref{fig:n_annuli} illustrates the general trend that a larger $N$ provides better accuracy at low noise levels up to a limit set by the beam size, but is less robust against noise, producing unconstrained fits or unphysical features with high noise. Empirically, we find that $N$ is limited by two conditions, (a) the mean annuli width, $r_\text{max}/N$, must be greater than half the FWHM of the beam and (b) the total S/N within a typical radial bin in the discretised 1D flux, $\vb* L$, should be greater than $\sim$~20. The optimal $N$ to use is the maximum $N$ that satisfies the conditions above. In practice, users are also encouraged to experiment with neighouring $N$ values to test the validity of the fit. 

\subsubsection{Runtime}
In our benchmark runtime tests performed on an Intel Core i7-8550U CPU with 1.80~GHz base frequency, we used images with dimensions of 200 by 200 pixels, PSF kernels with dimensions of 21 by 21 pixels and $N = 7$. Under these configurations, we estimate that each iteration of radial profile fitting takes approximately 0.2~s and each iteration of height fitting 2.0~s, if the rapid annuli method is used and all narrow annuli are cached. To generate and cache a set of rapid annuli (with 200 points per pixel when viewed face on) would require approximately 30~s. 

Convolution is a relatively computationally expensive step when using the rapid annuli generation method. The reason that height fitting is more computationally expensive is that for each individual height fit, the heights are fitted iteratively (unlike the radial profile matrix inversion), with each step requiring the re-generation of fitting annuli according to the updated heights at that step. 

In general, 100 iterations are required for the radial profile and height fitting algorithm to reach a reasonable precision, implying that the radial profile takes approximately 20~s to converge and the height 200~s under these configurations. 

\subsection{Summary of fitting procedure}
The fitting procedure can be summarised as follows:

\begin{enumerate}
\setcounter{enumi}{-1} 
\renewcommand{\labelenumi}{\arabic{enumi}}
    \item[(0)] Pre-processing: Rotate the disk major axis to be horizontal. Reshape to an even number of pixels and centre the image at the intersection of the four central pixels. Measure the background noise level. Apply the same processing to the PSF. 
    
    \item[(1)] Radial profile: Choose $N$ based on noise level and beam size. Perform automatic radial profile fitting for desired number of iterations. A face-on model image can be generated at this stage. 
    
    \item[(2)] Inclination: If there are no additional inclination constraints, perform automatic inclination fitting to obtain a lower bound for the inclination. 
    
    \item[(3)] Scale height: Choose region with non-negligible flux based on the radial profile fit. Perform automatic height fitting over that region with an inclination assumption. Repeat the fit for a range of possible inclinations based on available constraints. An edge-on model image can be generated, and structures in the residual image may be used to inform the more likely inclination and height combinations. 
\end{enumerate}

We also point out the following properties about the fitting method: 

\begin{itemize}
    \item The algorithm relies on the assumption that the disk is azimuthally symmetric, which can be verified if the recovered radial profile is consistent when fitted independently to the left and right sides of the disk. However, to improve the signal-to-noise ratio, the pre-processed imaged can be summed with its 180$^\circ$--rotated image about its centre before performing the fit. Correspondingly, the PSF kernel must also be summed with its rotated image and normalised. 
    
    %\item The radial profile fitting does not depend on knowledge about the inclination, and is therefore also valid for deconvolving images of nearly face-on disks. However, alternative methods that directly deconvolve the face-on, PSF-convolved flux profile are likely more useful.
    
    \item The primary merit of the directly fitted inclination constraints is to provide a more robust lower bound independent of the height fitting algorithm. However, in practice, when repeating the height fitting algorithm over the full range of allowed inclinations, one may encounter regions with negative height which are unphysical, thereby further tightening the inclination constraint. 
    
    \item The height fitting algorithm assumes that the height profile at each radial location is Gaussian by default. This is usually a reasonable approximation, however in a situation in which alternative distributions are more favourable, as has been suggested for the edge-on Beta Pic debris disk \citep{Matra2019} for example, this distribution can be readily modified. 
    
    \item Although taking the median of all individual radial profile and height fits---which are piecewise--constant---usually produces a relatively smooth curve, it may still contain a small amount of noise. By default, the algorithm smooths the curves with a Savitzky--Golay filter \citep{Savitzky1964} before plotting, which fits data points within a small moving window using polynomials, removing small and unphysical local oscillations without modifying the broader trend of the fitted surface brightness and height profiles. 
    
    \item The annuli generation method here uses the cylindrical-coordinate approximation rather than spherical coordinates. In reality, it may be more appropriate to consider material in debris disks as being ``shells'' rather than ``cylinders'', but at low scale heights this is a valid approximation. 
    
\end{itemize}

\section{Application to AU~Mic}
\label{sec:aumic}
AU~Mic is one of the nearest pre-main-sequence stars with an age of $\sim$~22~Myr \citep{Mamajek2014}. At a distance of $\sim$~9.7~pc \citep{Gaia2018}, AU~Mic hosts an extensively studied debris disk with well-resolved images in the optical \citep{Kalas2004, Krist2005}, near-IR \citep{Liu2004, Fitzgerald2007, Wang2015}, far-IR \citep{Matthews2015} and mm wavelengths \citep{MacGregor2013, Matthews2015, Daley2019}. Optical and far-IR imaging also revealed the presence of an extended halo surrounding the disk \citep{Matthews2015}. The system is also thought to host two close-in, Neptune-sized planets as suggested by TESS light curves \citep{Plavchan2020, Martioli2021}. 

% \subsection{Results}
We applied \texttt{Rave} to ALMA Band 6 $\sim$1.35 mm continuum observations of AU~Mic combined from three different epochs, March 2014, August 2014 and June 2015. A detailed description of the observational setup is available in \citet{Daley2019}. We retrieved the raw visibility datasets from the ALMA archive and calibrated them using the standard calibration scripts provided by the observatory, using the \textsc{CASA} software package \citep{McMullin2007}. AU~Mic exhibits significant stellar activity and is known to undergo strong episodic flares \citep{Smith2005}. We checked the visibility amplitudes as a function of time and manually flagged data in the time interval where the star flared strongly, corresponding to 24$^{\rm th}$ June 2015, between 04:23:38 and 04:27:30 \citep[in agreement with ][]{Daley2019}. 

The star exhibits significant proper motion over the course of the three observations, though ALMA accounts for this by maintaining the phase center at the proper motion-corrected stellar position, so that visibility phases for each dataset always refer to the stellar position appropriate for its observing date (and consequently images from each dataset are centred on the star). Nonetheless, to account for potential small astrometric offsets due to inaccurate phase referencing to the calibrator, we used a procedure analogous to \citet{Matra2020} and fitted the individual visibility datasets to obtain stellar offsets from the phase centre of each dataset separately. We then shifted the visibility datasets using these (small) best-fit offsets to align the datasets and ensure the star is exactly at phase centre of each observation. We then combined the three epochs to obtain a single, combined calibrated visibility dataset.

We imaged this calibrated dataset using the \textsc{tclean} task within CASA version 5.4.0. To image the continuum, we used multi-frequency synthesis mode and multiscale deconvolution. We used Briggs weighting with a robust parameter equal to 0.5 as a compromise between resolution and sensitivity, leading to an RMS noise level of 15 $\mu$Jy beam$^{-1}$ for beam size of 0.41"$\times$0.32". %Such variations present challenges to combining ALMA observations, and flares were accounted for before combining the data across the three epochs. 
Since the two sides of the disk showed no obvious asymmetry, we averaged the disk image (and beam) with its 180-degree rotation about its centre before performing any fitting to improve the signal-to-noise ratio. 

As the star is strongly detected, we included the stellar flux as a free parameter when performing the fits. The results are presented in Fig.~\ref{fig:aumic1}, \ref{fig:aumic2} and \ref{fig:aumic3}.

\subsection{Surface brightness}
\begin{figure*}
    \centering
    \includegraphics[width=18cm]{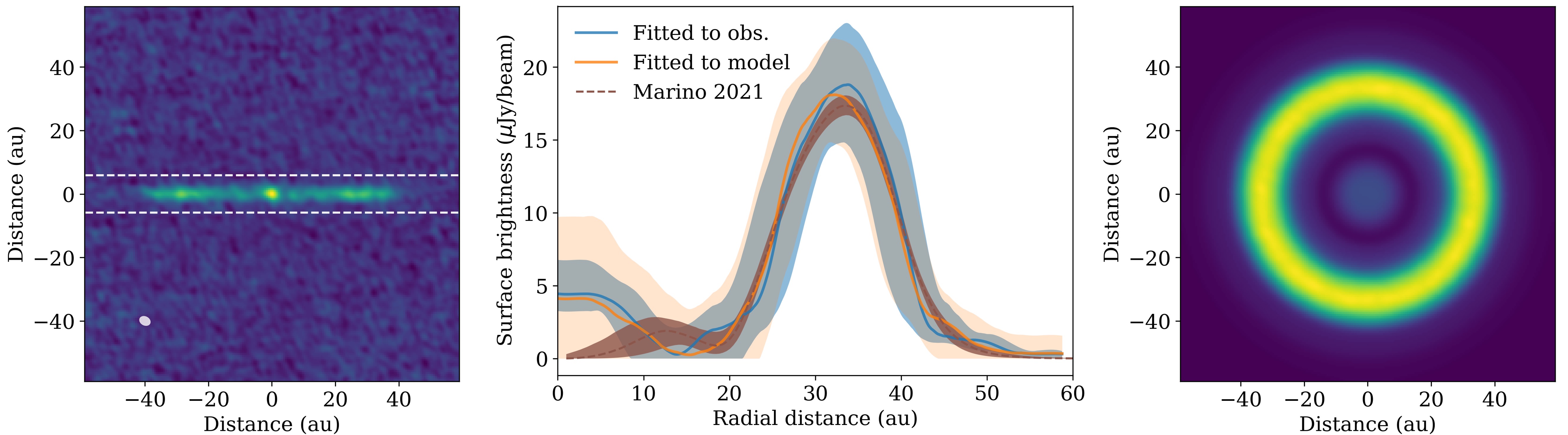}
    \caption{Left: Rotated ALMA Band 6 continuum image of AU~Mic combined from three different observing epochs. The beam is indicated in the bottom-left corner. Only the region within the dashed lines is used for fitting. Middle: Face-on surface brightness profile of AU~Mic recovered using \texttt{Rave} and the range of possible models fitted to the observation (shaded blue region) and median model (shaded orange region). The profile was jointly fitted with the stellar flux using $N = 7$. The stellar component is excluded from the plot displayed here. The profile recovered parametrically by \citet{Marino2021} is over-plotted with the 3$\sigma$ uncertainties in brown. Right: Face-on model image of AU~Mic using the best-fit surface brightness profile recovered in this work. }
    \label{fig:aumic1}
\end{figure*}

The surface brightness profile presented in Fig.~\ref{fig:aumic1} suggests that the observed flux originates from a disk extending from approximately 20 to 45~au and peaking at approximately 35~au. The inner and outer edges are smooth, resulting in an overall Gaussian-like profile. %, albeit slightly skewed towards the outer edge. 
Comparing this profile to that fitted by \citet{Marino2021} using parametric modelling with RADMC3D, the two fitted profiles are highly consistent: major features such as the rise, peak and fall of the two radial profiles are almost identical in steepness and location. 
%The flat-topped peak of our fit is not sufficiently significant to assert that it deviates radically from a Gaussian distribution, yet it is plausible that the distribution is more flat topped than implied by the fit in \citet{Marino2021}. 
However, the range of possible models of our fit suggests that a profile that deviates from a standard Gaussian (e.g., a sharper profile with a more flat-topped distribution) may also be consistent with the observations. 

Our fit also suggests a stellar flux value of 0.26~$\pm$~0.09~mJy. This is consistent with the value of 0.26~$\pm$~0.02~mJy fitted parametrically by \citet{Marino2021}, however it is significantly higher than the flux of approximately 0.05~mJy expected of the stellar photosphere as obtained from SED fitting (G. Kennedy, private communication). %In light of the variability of AU~Mic's stellar flux due to flares, such discrepancies are not unexpected.

After subtracting off the fitted central point source, the remainder of the radial profile still contains an extended central component (Fig.~\ref{fig:aumic1}). Integrating the best-fit profile between 0 and 15~au gives a total flux density of 0.09~mJy not accounted for by the fitted point source. This excess emission appears to support the possibility that there may be an additional component close to the star. 

The central peak was also detected by \citet{MacGregor2013}, in which the authors used a parametric model consisting of a central circular Gaussian component in addition to the main outer belt to fit the radial profile. Based on ALMA observations with lower resolution than in this study, the authors found that the central Gaussian component was best fit by a flux density of 0.32~$\pm$~0.06~mJy and standard deviation of $\leq$2.4~au. 

\citet{MacGregor2013} proposed that this possibly extended source of emission could be due to dust. Unusually high flux could be produced by stellar flares, but the chance of flares with this flux level is low. If this emission is indeed due to dust, this could imply the presence of an asteroid belt analogue \citep{MacGregor2013}. Higher resolution observations that target the inner 10~au region may be able to reveal its structure.

It is worth noting that \citet{Marino2021} fitted their profile directly in visibility space. The consistency between the two fits adds to evidence supporting the validity of fitting ALMA observations in image space (as opposed to visibility space) for the method developed here. 

As discussed in Sec.~\ref{sec:uncertainty}, the range of possible models associated with the non-parametric fit and uncertainty region of the parametric fit carry different meanings, and the parametric shaded region indicates the uncertainties under the assumption that the parametrisation used to perform the fit is the true parametrisation underlying the observation. Since it is not certain whether AU~Mic's radial profile actually takes on the assumed functional form, the plotted uncertainties may be an underestimate of the overall uncertainty on the radial profile. 

Our non-parametric fit suggests that the distribution may be smooth, in which case the range of possible models is broader than the likely distribution. However it does show the range of possible surface brightness values at each location, given that the distribution could contain discontinuities on scales of the beam size. 
This range of possible profiles therefore takes into account the broader range of shapes of the surface brightness profile that could feasibly reproduce the observed, projected 1D flux profile, and shall be considered a less biased estimate of the overall uncertainty.  %This is exemplified in Fig.~\ref{fig:radial}, in which the abrupt test case is not exactly reproduced by the best-fit profile, but is nonetheless mostly contained within the 1$\sigma$ region. 

\subsection{Scale height}
\begin{figure*}
    \centering
    \includegraphics[width=18cm]{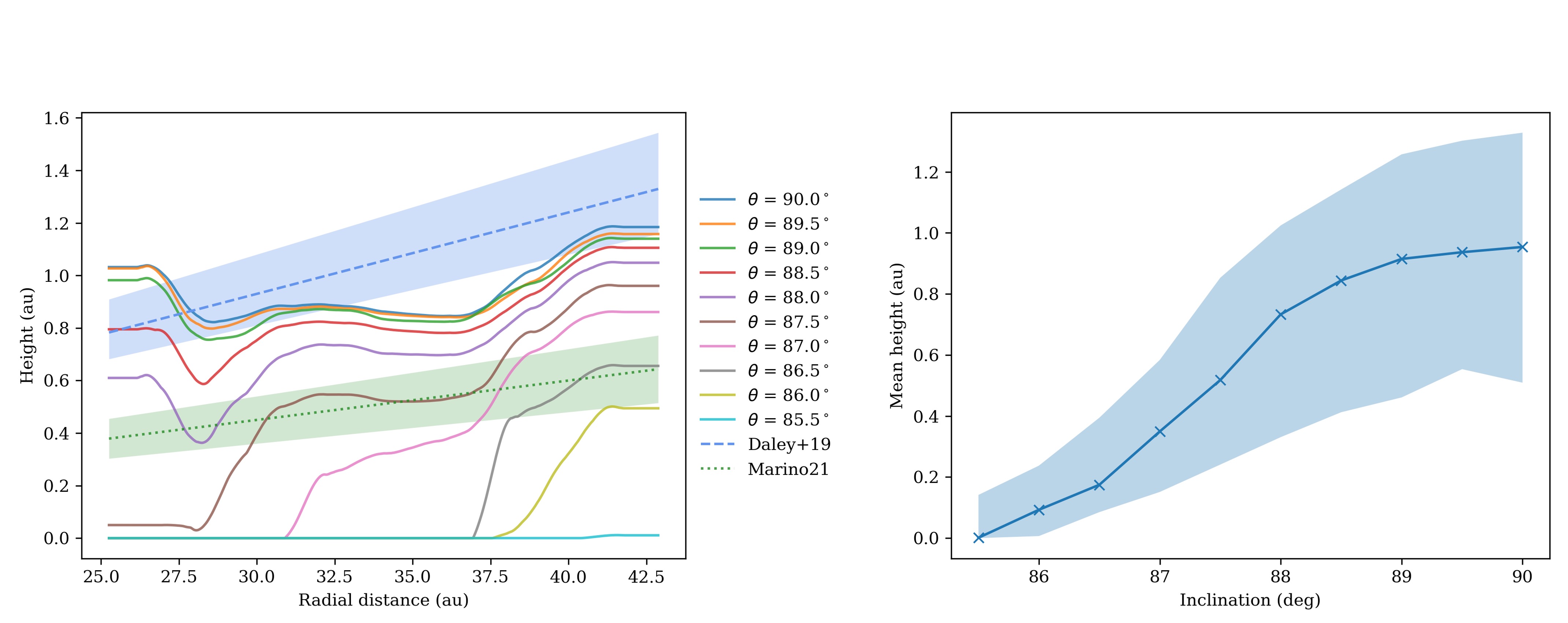}
    \caption{The left panel shows the best-fit scale height of AU~Mic under a range of inclination assumptions using \texttt{Rave}. The right panel shows the mean scale height of each curve as a function of inclination. 
    The region of the disk with substantial flux (as suggested by the fitted surface brightness profile) was fitted using $N = 3$. Regions where the surface brightness is consistent with 0 were not fitted to and were assumed to have a height of 1.2~au when fitting. Fig.~\ref{fig:aumic3} shows an example of the range of possible models associated with one inclination assumption. }
    \label{fig:aumic2}
\end{figure*}

\begin{figure*}
    \centering
    \includegraphics[width=18cm]{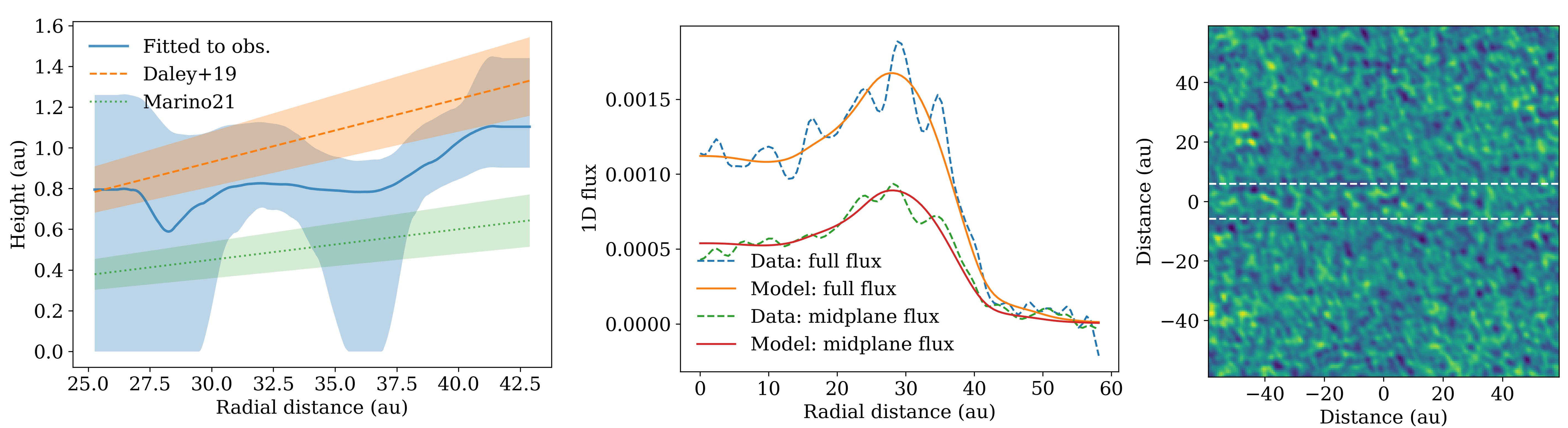}
    \caption{This figure shows the fitted scale height and range of possible models assuming an inclination of $88.5^\circ$. The full 1D flux and midplane flux of the observation and the best-fit model are displayed in the middle panel. The right panel shows the residual image obtained by subtracting the best-fit model from the data. A dashed box locates the approximate spatial extent of the disk. }
    \label{fig:aumic3}
\end{figure*}

Since there is no unique solution to the height profile due to its degeneracy with inclination, we present the height profiles under a range of inclination assumptions in Fig.~\ref{fig:aumic2}. Fig.~\ref{fig:aumic3} presents the outcome of the height fitting under one inclination assumption in more detail. 

As expected, the fitted height decreases as the assumed inclination decreases. At an inclination of $87^\circ$, the fitted height becomes 0 at some locations of the disk, implying that the inclination of AU~Mic cannot be lower than $87^\circ$. The example of the fit assuming an inclination of $88.5^\circ$ (Fig.~\ref{fig:aumic3}) adequately reproduced the observed midplane flux profile with no significant structures in the residual map. 

The $87^\circ$ lower bound for the inclination is consistent with a best-fit inclination of $88.5^\circ$ obtained by \citet{Daley2019}. \citet{Metchev2005} placed a lower bound of $87^\circ$ by modelling a perfectly flat disk and fitting to near-IR data taken by Keck/NIRC2, although their modelling prefers an inclination of $89^\circ$.

\citet{Daley2019} fitted to the height profile of AU~Mic assuming a constant aspect ratio, i.e. $h/r$ value, across the disk. Their best-fit model has an inclination of $88.5^\circ$ and a global aspect ratio of $0.031^{+0.005}_{-0.004}$, which translates to approximately 0.8~au at $r = 25$~au and 1.2~au at $r = 40$~au (see Fig.~\ref{fig:aumic2} for comparison). Using similar parametric modelling approaches, \citet{Marino2021} obtained a best-fit model with an inclination of $88.2^\circ$ and an aspect ratio of $0.015^{+0.003}_{-0.003}$, which is significantly lower than that obtained by \citet{Daley2019} (despite the slightly lower inclination in \citealt{Marino2021}). 

Scale height measurements can be linked directly to the level of dynamical stirring. Using their derived $h/r$ value, \citet{Daley2019} constrained any stirring bodies in the disk to be greater than approximately 400~km in radius, corresponding to about 0.05 Pluto masses, but less than approximately 1.8 Earth masses to produce the observed disk thickness, provided there are no external stirring bodies. The authors used this result to rule out the presence of Neptune-sized bodies in the debris disk. 

Our $\theta=88.5^\circ$ height profile is marginally lower than that obtained by \citet{Daley2019}, but is sightly higher than that obtained by \citet{Marino2021}. Our fitted profile appears to favour a nearly constant height across the main flux-emitting region of the disk, although a height profile proportional to the radius cannot be rejected. Any future observational constraints on the inclination will be able to lift the degeneracy between height and inclination, thereby informing the more likely height profile among those presented in Fig.~\ref{fig:aumic2}.

\section{Conclusions}
We presented a non-parametric method to recover the radial surface brightness profile and scale height profile of edge-on debris disks, taking as input an image of a disk and the PSF of the corresponding observation. Our method operates under the assumptions that (1) the disk is azimuthally symmetric, (2) the disk is optically thin and (3) the vertical distribution of material at any radial location can be approximated by a Gaussian. 

Non-parametric fitting has the advantage over parametric fitting that no functional form for the surface brightness and scale height profiles is assumed, thereby removing biases towards a particular shape or form and providing more realistic estimates of the range of possible radial and height profiles. This is particularly useful in the context of searching for substructures, which provide the best constraints for dynamical interactions with planets that are otherwise difficult to detect. Our method would then inform more realistic constraints on any potential planets in the system, the parameters of which should ideally be based on an understanding of the radial and height profiles of the disk and their uncertainties that are independent of the choice of parametrisation. In many cases, non-parametric modelling may suggest a larger range of possible models compared to the uncertainties under parametric fitting, but observations with higher sensitivity and resolution can tighten these constraints. 

We demonstrated with a series of test cases that the method is able to converge to a range of radial and height profiles within uncertainties. While the radial profile fit is unique, the scale height profile is degenerate with inclination. We developed a method to independently set constraints for the inclination. Within the plausible inclination range, the height profile may be fitted over a range of inclination assumptions. The residual map may further inform the more likely inclination and height combination. 

Applying this method to ALMA observations of the AU~Mic debris disk, we recover a surface brightness profile consistent with that obtained by \citet{Marino2021} using parametric modelling, but with uncertainties that are unbiased by assumptions of the parametrisation. We also obtain the scale height of AU~Mic as a function of radial location and quantify its degeneracy with inclination. Our scale height fitted at 88.5$^\circ$ inclination is broadly consistent with the global aspect ratio ($h/r$) obtained by \citet{Daley2019} with parametric modelling, but suggests that a flat profile with a height of $\sim$0.8 au across the whole disk is also possible. 

The algorithms are implemented as an open-source package called \texttt{Rave} in the \texttt{Python} language and is available for download along with examples. As more high-resolution observations of edge-on disks are made, we expect that \texttt{Rave} will be useful to the community to complement parametric methods to elucidate the structure of highly inclined debris disks. 

%%%%%%%%%%%%%%%%%%%%%%%%%%%%%%%%%%%%%%%%%%%%%%%%%%
\section*{Acknowledgements}
The authors acknowledge preliminary work on the method presented here as part of a Part III project at the University of Cambridge by D. Payne. 
Y.H. acknowledges funding from the Gates Cambridge Trust. L.M. acknowledges funding from the European Union’s Horizon 2020 research and innovation programme under the Marie Sklodowska-Curie grant agreement No 101031685. This research made use of NASA's Astrophysics Data System; the \textsc{IPython} package \citep{ipython}; \textsc{NumPy} \citep{numpy}; \textsc{matplotlib} \citep{matplotlib}; and \textsc{Astropy}, a community-developed core Python package for Astronomy \citep{astropy}.

\section*{Data Availability}
The data used in this study are available on the ALMA archive under program IDs 2011.0.00142.S, 2012.1.00198.S and 2016.1.00878.S.

\bibliographystyle{mnras}
\bibliography{references}

%%%%%%%%%%%%%%%%% APPENDICES %%%%%%%%%%%%%%%%%%%%%
\appendix

%\section{Appendix}

%%%%%%%%%%%%%%%%%%%%%%%%%%%%%%%%%%%%%%%%%%%%%%%%%%
\bsp
\label{lastpage}
\end{document}